\documentclass[11pt,preprintnumbers]{article}
\usepackage{jheppub}
\pdfoutput=1
\usepackage{bbm}
\usepackage{epsfig,bm}
\usepackage{graphics,graphicx,comment,subfigure,caption}
\usepackage{amsmath}
\usepackage{mdframed} 
\usepackage{epigraph}
\usepackage{empheq}
\usepackage{framed}
\usepackage{slashed}
\usepackage{mathtools}

\usepackage{color}%

\def\bea{\begin{eqnarray}}
\def\ea{\end{eqnarray}}
\def\Th#1#2{\vartheta{\tiny\begin{bmatrix}
{#1}\\
{#2}
\end{bmatrix}}}

\newcommand{\tr}{\textrm{Tr}\,}

\newcommand{\im}{\mathrm{Im}\,}
\newcommand{\re}{\mathrm{Re}\,}

\def\half{{\textstyle{1\over 2}}}

\hypersetup{
  colorlinks=true,    % false: boxed links; true: colored links
  linkcolor=blue,     % color of internal links (change box color with linkbordercolor)
  citecolor=blue,    % color of links to bibliography
  filecolor=magenta,   % color of file links
  urlcolor=blue      % color of external links
}

\begin{document}

\title{Modularity and 4D-2D spectral equivalences for large-$N$ gauge theories with adjoint matter}

\author[1]{G\"{o}k\c{c}e Ba\c{s}ar,}
\emailAdd{gbasar@umd.edu} 
\author[2]{Aleksey Cherman,}
\emailAdd{aleksey.cherman.physics@gmail.com} 
\author[3,1]{Keith R. Dienes}
\emailAdd{dienes@email.arizona.edu} 
\author[4, 5]{and David A. McGady}
\emailAdd{mcgady@nbi.ku.dk} 
\affiliation[1]{Department of Physics, University of Maryland, College Park, MD 20742  USA}  
\affiliation[2]{Institute for Nuclear Theory, University of Washington, Seattle, WA  98195 USA}
\affiliation[3]{Department of Physics, University of Arizona, Tucson, AZ 85721 USA}
\affiliation[4]{Kavli IPMU, University of Tokyo, Kashiwa, Chiba 277-8583 Japan}
\affiliation[5]{Niels Bohr International Academy \& Discovery Center, Niels Bohr Institute, \\ University of Copenhagen, 2100 Copenhagen, Denmark}

\abstract{
In recent work, we demonstrated that the confined-phase spectrum of non-supersymmetric pure Yang-Mills theory coincides with the spectrum of the chiral sector of a two-dimensional conformal field theory in the large-$N$ limit.  This was done within the tractable setting in which the gauge theory is compactified on a three-sphere whose radius is small compared to the strong length scale.  In this paper, we generalize these observations by demonstrating that similar results continue to hold even when massless adjoint matter fields are introduced.  These results hold for both thermal and $(-1)^F$-twisted partition functions, and collectively suggest that the spectra of large-$N$ confining gauge theories are organized by the symmetries of two-dimensional conformal field theories.
}

\maketitle

%%%%%%%%%%%%%%
\section{Introduction}
\label{sec:Introduction}
%%%%%%%%%%%%%%
In the large-$N$ limit, QCD and other 4D confining gauge theories become free in terms of their physical degrees of freedom~\cite{tHooft:1973jz,Witten:1979kh}. The first step towards a solution of a confining large-$N$ theory entails determining which particular free theory it becomes at large $N$ by specifying the spectrum of particle masses. This amounts to determining the two-point functions of the theory. Once this is done, one would then want to characterize the large-$N$ limit of the connected correlation functions of three or more operators. Progress towards these goals has been made for situations in which these gauge theories are supersymmetric; for a review see Ref.~\cite{Beisert:2010jr}. Unfortunately, there has been much less progress for more realistic theories that lack supersymmetry.  Indeed, for non-supersymmetric confining 4D gauge theories, such as QCD, even the first step of determining the large-$N$ particle mass spectrum has thus far been beyond reach. 
 
In recent work~\cite{Basar:2015xda}, we focused on the case of pure, non-supersymmetric, Yang-Mills (YM) theory (i.e., Yang-Mills theory without matter fields) and demonstrated that its confined-phase spectrum coincides with the spectrum of the chiral sector of a two-dimensional conformal field theory (CFT) in the large-$N$ limit.  This was done at finite temperature $\beta\equiv 1/T$, within the tractable setting in which the gauge theory is compactified on a round three-sphere whose radius is small compared to the strong length scale.  

\begin{figure}[t]
\centering
\includegraphics[width=0.7\textwidth]{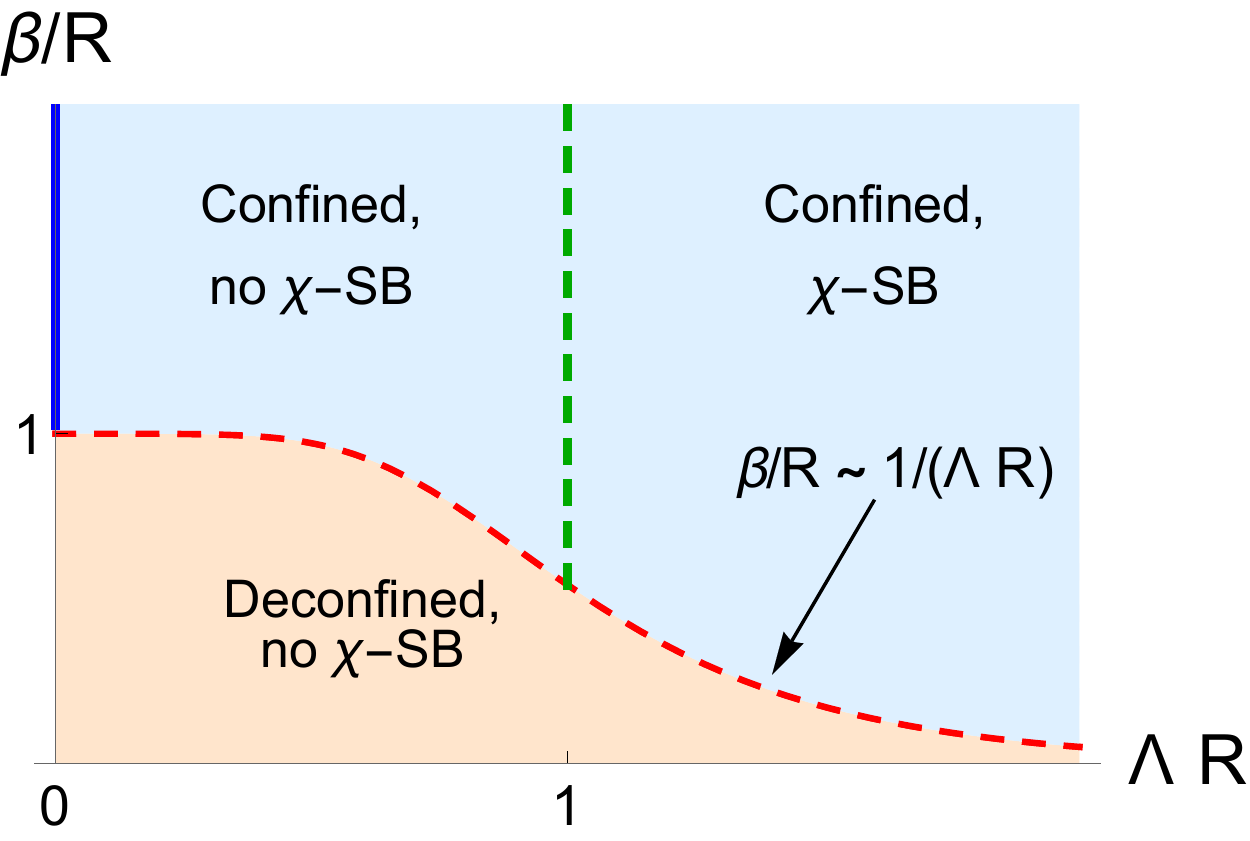}
\caption{A conjectured phase diagram for large-$N$ gauge theories compactified on $S^3_{R} \times S^1_{\beta}$. The dashed red curve indicates a phase transition to the deconfined phase.  At small $R\Lambda$, it can be shown that the deconfinement transition takes place when $R\Lambda \sim 1$.  For theories that have a mass gap $\sim \Lambda$ in the $R\Lambda \gg 1$ limit, one would expect a deconfinement transition at $\beta \sim 1/\Lambda$.  The curve sketched in the diagram is the simplest interpolation between these two limiting behaviors.  The dashed green line indicates a possible chiral symmetry-breaking ($\chi$-SB) phase transition.  As emphasized in Ref.~\cite{Basar:2014jua}, these phase transitions may or may not be present, depending on the matter content and the boundary conditions for the fermions.  The blue line on the left edge indicates the region for which we find a 2D description of the 4D theory.}
\label{fig:QCDPhaseDiagram}
\end{figure}

In this paper, we shall generalize the analysis of Ref.~\cite{Basar:2015xda} to the broader case of asymptotically-free gauge theories with $n_f$ massless Majorana adjoint fermions and $n_s$ massless (conformally-coupled) adjoint scalars. Just as in Ref.~\cite{Basar:2015xda}, we shall consider this theory compactified on a round three-sphere $S^3$ with radius $R$ and we shall work in the $R\Lambda\to 0$ limit, where $\Lambda$ is the strong scale associated with the gauge theory. This limit is particularly attractive because as $R\Lambda$ becomes small, the 't~Hooft coupling $\lambda$ at the scale $R$ approaches zero.  As a result, these theories can be solved in the $R\Lambda \to 0$ limit. Moreover, at large $N$, adjoint-matter gauge theories can be shown to be in a confining phase even when $R\Lambda \to 0$. Here confinement is defined to be associated with an unbroken center symmetry and a free energy that scales as $N^0$, as discussed in Ref.~\cite{Aharony:2005bq}. In particular, there are known closed-form expressions for the large-$N$ confining-phase thermal partition functions when $\lambda=0$ (see, e.g., Refs.~\cite{Aharony:2003sx,Sundborg:1999ue,Polyakov:2001af}). A conjectured phase diagram for this class of theories is sketched in Fig.~\ref{fig:QCDPhaseDiagram}. 

%%%%%%%%%%%%%%
%%%%%%%%%%%%%% 

Understanding the symmetry structure of the spectrum in the solvable $R\Lambda \to 0$ corner of the phase diagram of adjoint-matter confining gauge theories is likely to be a valuable and perhaps necessary step toward understanding the structure of the spectrum of confining gauge theories for more general $R\Lambda$. Understanding this structure is therefore the main thrust of this paper. Quite remarkably, although the analysis of Ref.~\cite{Basar:2015xda} was limited to pure Yang-Mills theory, in this paper we find that similar results continue to hold even when massless adjoint matter fields are introduced. Specifically, we find that the confined-phase spectra of large-$N$ four-dimensional quantum field theories (QFTs) on $S^3\times S^1$ are identical to the spectra of certain two-dimensional (2D) CFTs in the regime described above.  More precisely, at large $N$, the $S^3 \times S^1$ partition functions $Z_{\rm 4D}$ coincide with certain chiral torus partition functions $Z_{\rm 2D}$ of 2D CFTs, so that we obtain a relation of the form 
\begin{align}
Z_{\rm 4D}(\tau) &= Z_{\rm 2D}(\tau).
 \label{eq:TheClaim}
\end{align}
In writing this result, we have taken advantage of the fact that the functions $Z_{\rm 4D}$ are meromorphic functions of $\beta/R$ in order to analytically continue $\beta/R$ into the complex plane, setting $\beta/R = 2\pi i \tau$ where $\tau$ is generally complex. Here $\im \tau =\beta/(2\pi R) = C_{S^1}/C_{S^3}$ is the ratio of the circumferences of $S^1$ and $S^3$. On the 2D side of the relation, $\im \tau$ is the ratio of the cycles of a torus, as usual,  while $\re \tau$ controls the momentum on the spatial cycle. The physical meaning of $\re \tau$ on the 4D side of the relation is in general less evident. For 4D theories with fermions, we will see that the modular $T$-transformation $\tau \to \tau+1$ (which generates non-zero integer values of $\re \tau$) has the effect of flipping the fermion boundary conditions on $S^1$ from periodic to anti-periodic.  We leave the interesting and important challenge of understanding the physical meaning of generic points along the ($\re \tau$)-direction to future work.

The result in Eq.~\eqref{eq:TheClaim} is interesting from the perspective of the general goal of understanding the structure of the large-$N$ spectrum. Recall that the definition of a generic free QFT  relies on a large set of parameters whose number scales with the number of distinct single-particle  excitations of the QFT. However, the number of parameters is reduced in the presence of symmetries. The spectrum of a given large-$N$ confining gauge theory consists of an infinite number of single-particle excitations, even in the $R\Lambda \to 0$ limit, but such theories have very few adjustable parameters. For instance, pure $SU(N)$ Yang-Mills theory has no dimensionless parameters at all in the large-$N$ limit, both in the $R\Lambda \to \infty$ limit and in the $R\Lambda \to 0$ limit. It is therefore tempting to wonder whether the large-$N$ spectrum is controlled by some emergent spectrum-generating symmetry. Of course, even if such symmetries exist at large $N$, presumably they are broken at finite $N$, and they may not be apparent in a Lagrangian description of the theory based on the microscopic quark and gluon fields.  It is not currently clear how to explore the structure of the confined-phase spectrum for generic $R\Lambda$, but in the $R\Lambda \to 0$ limit the problem simplifies dramatically since the spectrum in that limit is known.  What our result in  Eq.~\eqref{eq:TheClaim} suggests is that the $R\Lambda \to 0$ spectrum is controlled by the symmetries of a 2D CFT.~  Moreover, such 2D CFTs are known to have infinite-dimensional symmetries, because their spectrum-generating symmetry algebras always include at least one copy of the infinite-dimensional Virasoro symmetry.  Our observations thus suggest that the large-N confined-phase spectra of 4D gauge theories are controlled by infinite-dimensional spectrum-generating algebras which include at least the Virasoro algebra, at least in the small $R\Lambda$ limit.  It would be very interesting to understand to what extent this generalizes for generic $R\Lambda$.

String theory provides additional reasons to suspect a connection between 2D CFTs and 4D gauge theories. Large-$N$ confining gauge theories are believed to be describable as free string theories, and free string theories have a world-sheet description as 2D CFTs.  However, as we shall discuss in the conclusions, our results do not fit easily with such string worldsheet-based expectations. Understanding the string-theoretic underpinnings of our results therefore remains an exciting open question.

A relation such as that in Eq.~\eqref{eq:TheClaim} may seem surprising for many reasons.  At the most basic level, it may seem implausible that the partition functions of QFTs defined  in different numbers of spacetime dimensions could possibly be identical.  Indeed, the result in Eq.~\eqref{eq:TheClaim} might initially appear to be inconsistent with the properties of typical 4D QFTs, because such theories typically exhibit the asymptotic behavior
\begin{align}
\lim_{\beta\to 0} Z^{\rm generic}_{\rm 4D}(\beta) ~\sim~  e^{- \frac{\sigma_{4} R^3}{\beta^3}} .
\label{eq:4Dbehavior}
\end{align}
By contrast, for a 2D CFT one instead expects
\begin{align}
\lim_{\beta\to 0}  Z_{\rm 2D}(\beta) ~\sim~ e^{- \frac{\sigma_{2} R}{\beta}}.
\label{eq:2DBehavior}
\end{align}
This latter behavior can be understood from the observation that the partition functions of 2D CFTs have simple properties under modular transformations acting on $\tau$, and thus are expected to be expressible as combinations of modular forms and Jacobi forms which are functions of $\tau$. The modular properties of such functions then lead to the limiting behavior in Eq.~\eqref{eq:2DBehavior}.

In general, for 4D theories we would expect to observe the behavior in Eq.~\eqref{eq:4Dbehavior}, and so we would not expect 4D QFT partition functions to be expressible as finite products of modular forms.  However, large-$N$ confining gauge theories are very special 4D QFTs.  As discussed in Refs.~\cite{Basar:2014hda,Basar:2014jua}, there exists numerical evidence that the large-$N$ confined-phase partition functions discussed above scale as in Eq.~\eqref{eq:2DBehavior} for small $\beta$, rather than as in Eq.~\eqref{eq:4Dbehavior} --- as long as the $|\tau| \to 0$ limit is taken before the $\arg \tau \to \pi/2$ limit, i.e., as long as $\beta \sim i \tau\to 0$  along a contour that is slightly off the real-$\beta$ axis. The ordering of limits can be important due to Hagedorn singularities. 

As we shall demonstrate in this paper, the results obtained in Refs.~\cite{Basar:2014hda,Basar:2014jua} are possible 
because the large-$N$ confined-phase partition functions of gauge theories on $S^3 \times S^1$ can indeed be expressed as combinations of modular and Jacobi forms. This surprising ``modularity'' is thus an important ingredient governing the spectra of such theories, and enables these 4D partition functions to resemble the chiral torus partition functions of 2D CFTs, as claimed in Eq.~\eqref{eq:TheClaim}. Thus, in this sense, the results in this paper both confirm and extend those of Refs.~\cite{Basar:2014hda,Basar:2015xda,Basar:2014jua}.  Furthermore, as we shall see, they even allow us to extract some of the properties of the 2D CFTs to which our 4D gauge theories are isospectral.

This paper is organized as follows. In Sect.~\ref{sec:PartitionFunctions} we begin by discussing the calculation of the large-$N$ partition functions of the 4D theories which are our main focus in this paper. Then, in Sect.~\ref{sec:ModularForms} we discuss the modularity properties of these 4D partition functions. In Sect.~\ref{sec:Implications} we discuss the various physical features that flow directly from this modularity and demonstrate that the large-$N$ 4D gauge theory partition functions can be written as the partition functions of 2D CFTs.  In Sect.~\ref{sec:ModularCompletion} we explore some properties of these  2D CFTs. Finally, in Sect.~\ref{sec:Trees} we conclude by listing a number of open questions and discussing how our results relate to previous observations in the existing literature. Several appendices are also included which define the notation and conventions that we shall be using throughout this paper and which provide further details concerning some of the results derived.

%%%%%%%%%%%%%%%%%%%%
\section{Calculation of large-$N$ partition functions}
\label{sec:PartitionFunctions}
%%%%%%%%%%%%%%%%%%%%

In this section we review the construction of large-$N$ confining-phase partition functions on $S^3 \times S^1$. 
%%%%%%%%%%%
\subsection{Large-$N$ limit and compactification on $S^3 \times S^1$}
\label{sec:LargeN}
%%%%%%%%%%%
We work in the 't~Hooft large-$N$ limit, with $N \to \infty$ while all other scales are held fixed. In asymptotically-free 4D gauge theories with gauge coupling $g$, the one-loop relation between the strong scale $\Lambda$ and a UV cutoff scale $\mu_{\rm uv} $ is
\begin{align}
\Lambda = \mu_{\rm uv} \, e^{-\frac{8\pi^2}{\beta_0 \lambda(\mu_{\rm uv})}}
\end{align}
where $\beta_0$ is the one-loop coefficient of the $\beta$-function for $\lambda(\mu_{\rm uv} ) = g^2(\mu_{\rm uv} ) N$, normalized such that $\beta_0 = 11/3$ in $SU(N)$ Yang-Mills theory. In the 't~Hooft large-$N$ limit one wishes to keep $\Lambda$ independent of $N$. To this end, one sets  $\lambda(\mu_{\rm uv} )$ and $\mu_{\rm uv} $ to be $N$-independent. We assume that $n_f, n_s$ are independent of $N$, and also take $R$ and $\beta$ to be independent of $N$. Then planar Feynman diagrams dominate at large $N$ and the standard $N$-counting rules follow. As is common in studies of large-$N$ theories, we focus on the $U(N)$ theories when discussing the $N \to \infty$ limit.\footnote{The overall $U(1)$ completely decouples for any $N \ge 1$ in adjoint-matter theories, even at finite 't~Hooft coupling, so its contribution to the partition function factorizes and could easily be taken into account if one wanted to write down results for the $N \to \infty$ limit of $SU(N)$ theories.}

When $R\Lambda \to 0$, the asymptotically-free gauge theories we consider become essentially free. A quick way to see this is that if $R\Lambda \ll 1$, the relevant scale for the 't~Hooft coupling becomes $1/R$, and $\lambda(1/R) \to 0$ thanks to asymptotic freedom. We work to leading order in the small $R\Lambda \to 0$ limit, which amounts to taking $\lambda = 0$. The phase diagram of the theory as a function of $R\Lambda$ and $\beta/R$ is sketched in Fig.~\ref{fig:QCDPhaseDiagram}.

%%%%%%%%%%%
\subsection{Derivation of thermal and $(-1)^F$-twisted partition functions}
\label{firstZcomp}
%%%%%%%%%%%
We now review the computation of the thermal and $(-1)^F$-twisted partition functions for large-$N$ gauge theories with adjoint matter on $S^3 \times S^1$.  These partition functions 
are respectively defined as 
\begin{align}
Z(\beta) &= \tr e^{-\beta H} \nonumber\\
\tilde{Z}(\beta) &= \tr (-1)^{F} e^{-\beta H}.
\end{align}
At large $N$, the computation of $Z(\beta)$ and $\tilde{Z}(\beta)$ can be organized into three steps:
\begin{enumerate}
\item Construct partition functions counting single-particle excitations of the gluon and matter fields. This comprises the set of operators that can be inserted into single- and multi-trace operators in the full theory.
\item Construct the partition function for the physical single-particle excitations of the large-$N$ gauge theory. This corresponds to specifying the energies and degeneracies of all single-trace operators in the theory.
\item Construct the full grand-canonical partition functions $Z(\beta)$ and $\tilde{Z}(\beta)$, which count all the physical multi-particle excitations as well as single-particle excitations.
\end{enumerate}
In what follows we briefly summarize each of these steps, with an emphasis on the issues which will be important for the rest of our analysis.

First, we discuss the partition functions for the excitations of the fundamental gauge and matter fields. In the weakly-coupled $R\Lambda \ll 1$ limit, the microscopic fields of the gauge theory --- the gluon and matter fields --- can be represented as infinite collections of harmonic oscillators, all with non-vanishing oscillation frequencies set in units of $1/R$. There are three types of harmonic oscillator fields that we can include, associated with microscopic scalar, fermion, and massless vector fields. The energies and degeneracies of the operators associated to these fundamental fields are counted by the so-called `letter' partition functions $z_{s}, z_{f}, z_{v}$ respectively, which can be written as
\begin{align}
z_s(q)& = {q^{1/2}+q^{-1/2}\over(q^{-1/2}-q^{1/2})^3} = \frac{q + q^2}{(1-q)^3}~\nonumber\\  
z_f(q)& = {4\over(q^{-1/2}-q^{1/2})^3} = 4\frac{q^{3/2}}{(1-q)^3} \,  \label{eq:Lynchpin} \nonumber\\ 
1-z_v(q)& = \frac{(q^{3/2}+q^{-3/2})-3(q^{1/2}+q^{-1/2})}{(q^{-1/2}-q^{1/2})^3} = \frac{(1+q^3)-3(q +q^2)}{(1-q)^3} \,  
\end{align}
where we have defined $q \equiv \exp\left(-{ \beta\over R} \right)$. Thus $z_s$, $z_f$, and $z_v$ are real-analytic functions of $\beta$, and the states of the adjoint-matter gauge theory are built from combinations of these microscopic fields. 

For what follows, it will be important to remember where these expressions come from. As discussed, e.g., in Ref.~\cite{Aharony:2003sx}, a free conformally-coupled massless scalar field on $S^3_{R} \times S^1_{\beta}$  has single-particle excitation energies that can be written as $E_n = (n+1)/R$, $n=0,1,2, \ldots$, with degeneracies $d_n = (n+1)^2$. The associated single-particle partition function is
\begin{align}
z_s(\beta) &= \sum_{n=0}^{\infty} (n+1)^2 e^{-\beta (n+1)/R}e^{-\beta \epsilon/R}\nonumber\\
&= q^{\epsilon} \frac{q^{-1/2}+q^{1/2}}{(q^{-1/2}-q^{1/2})^3}
\end{align}
where $\epsilon$ parametrizes the {\it a priori} arbitrary choice of vacuum energy. Very similar calculations yield $z_f$ and $z_v$. 

We now make the algebraic observation that if (and only if) we set $\epsilon = 0$, the single-letter partition functions $1-z_{v}(q), z_{f}(q), z_{s}(q)$ transform to $-[1-z_{v}(q)], -z_{f}(q),- z_{s}(q)$ under the formal `T-reflection' operation $\beta \to -\beta$ (i.e., $q \to 1/q$ and $q^{1/2} \to q^{-1/2}$).  Indeed, this observation was a key step in the demonstration of a more subtle temperature-reflection symmetry of $Z(\beta)$ and $\tilde{Z}(\beta)$ in Ref.~\cite{Basar:2014mha}, under which these grand-canonical partition functions transform into themselves up to a temperature-independent phase.  This choice for $\epsilon$ was made in writing Eq.~\eqref{eq:Lynchpin}, and the single-letter partition functions in Eq.~\eqref{eq:Lynchpin} are written in two different ways to emphasize their T-reflection properties.  These will be important in our analysis of modularity properties of $Z$ and $\tilde{Z}$ below.

Now let us consider the physical single-particle excitations. The spectral problem in the weakly-coupled gauge theory remains somewhat non-trivial due to the color Gauss-law constraint, which is present for any non-zero $\lambda$, no matter how small. The Gauss law implies that the physical states are created by single and multi-color-trace operators hitting the vacuum. This must be taken into account if we wish the $\lambda = 0$ theory to describe a limit of a theory with $\lambda \to 0^{+}$. Thus, in order to compute the spectrum of a non-Abelian gauge theory, we must count the energies and degeneracies of collections of harmonic oscillators drawn from $z_s, z_{f}, z_{v}$, subject to the color-singlet constraint.

At large $N$ and in the confining phase, the single-particle states are single-trace states while multi-trace states are multi-particle states. Taking the 't~Hooft large-$N$ limit defined above sharpens the distinction between single-trace and multi-trace operators and dramatically simplifies the counting problem yielding the partition function. If we were to work in a non-'t~Hooft large-$N$ limit and were to consider the contributions of states with energies that scale with $N$, then there would be algebraic relations between states created by single-trace chains of $N$ operators and multi-trace operators.  The counting problem would then be difficult. Fortunately, our assumption that the cutoff $\mu_{\rm uv}$ scales as $\mu_{\rm uv} \sim N^0$ means that we only need to consider states with energies $\sim N^0$.

The physical single-particle partition functions are just the single-trace partition functions, which turn out to be~\cite{Aharony:2005bq,Polyakov:2001af,Sundborg:1999ue}
\begin{align}
Z_{\rm ST} &= -\sum_{k=1}^{\infty} \frac{\varphi(k)}{k} \log \left[1 - z_v(q^{k})+(-1)^{k} n_f z_{f}(q^k)- n_s z_{s}(q^{k})\right] \nonumber\\
\tilde{Z}_{\rm ST} &=  -\sum_{k=1}^{\infty} \frac{\varphi(k)}{k} \log \left[1 - z_v(q^{k})+ n_f z_{f}(q^k)- n_s z_{s}(q^{k})\right] .
\end{align}
Here $\varphi(k)$ is the Euler totient function, which counts the integers smaller than $k$ which are coprime to $k$. These expressions are built to correctly encode the cyclic permutation properties of single-trace operators, with attention to the combinatorics of repeated operators.

We can now write down the full grand-canonical partition functions. At large $N$ the single-trace states do not interact, and the space of multi-trace states is a Fock space built out of single-trace states. As a result, the grand-canonical and single-particle partition functions are related through the plethystic exponential, $Z(q) = \exp\left[\sum_{n=1}^{\infty}\frac{1}{n}Z_{\rm ST}(q^n)\right]$. One can show that the grand-canonical partition functions take an even simpler form than the single-trace partitition functions~\cite{Aharony:2005bq,Polyakov:2001af,Sundborg:1999ue}:
\begin{align}
Z(q; n_s,n_f) &= \tr e^{-\beta H} = \prod_{n=1}^\infty {1\over 1-z_v(q^{n})+(-1)^n n_f\, z_f(q^{n})-n_s\,z_s(q^{n})}\,\nonumber\\
\tilde{Z}(q; n_s,n_f)&= \tr (-1)^F e^{-\beta H} = \prod_{n=1}^\infty {1\over 1-z_v(q^{n})+n_f\, z_f(q^{n})-n_s\,z_s(q^{n})}\,.
\label{eq:Zt}
\end{align}

The partition functions in Eq.~\eqref{eq:Zt} are infinite products of rational functions in $q = e^{-\beta/R}$, and $q$ is a real-analytic function of $\beta/R$. If we analytically continue $\beta/R \in \mathbb{R}^{+}$ to a complex parameter $\beta/R \to -2\pi i \tau$ with $\tau = t_1 + i t_2$, so that $t_2 = \beta/(2\pi R)$, the confining-phase partition functions become meromorphic functions of $\tau \in \mathbb{H}$, the complex upper half-plane. In this paper, we shall show that $Z(\tau)$ and $\tilde{Z}(\tau)$ are built out of modular forms and Jacobi forms with modular parameter $\tau$, and explore the consequences of this fact. Indeed, we shall see that these observations hold for all $n_f$ and $n_s$.

%%%%%%%%%
\subsection{Comments on confinement in the small-$R\Lambda$ limit}
%%%%%%%%%
Adjoint-matter gauge theories in the limit relevant to Eq.~\eqref{eq:Zt} behave in the ways that one would expect from well-to-do confined-phase gauge theories, at least as long as $\beta \gtrsim R$ ~\cite{Aharony:2003sx,Unsal:2007fb}:
\begin{itemize}
\item The thermal and twisted free energies scale as $N^0$. 
\item Center symmetry is unbroken.
\end{itemize}
We note that the realization of center symmetry and the large-$N$ scaling of the free energy are the only two commonly-used order parameters for confinement at large $N$ that make sense within finite volumes. Some other popular order parameters, such as the string tension inferred from the energy of a pair of heavy probe quarks as they become widely separated, must be defined in an infinite-volume limit.  Thus, given that our goal is to use $R\Lambda$ as a control parameter for the study of the large-$N$ confined-phase spectrum, it seems reasonable to characterize confinement by these two order parameters.

As a consequence of their unbroken center symmetry, gauge theories on $S_R^3 \times S_{\beta}^1$ enjoy large-$N$ volume independence in the size of $S^1$~\cite{Unsal:2007fb}. Also, the thermal densities of states $\rho(E)$  have a Hagedorn behavior $\rho(E) \to e^{+\beta_H E}$ for large $E$ in the confined phase. (In Ref.~\cite{Cohen:2015hwa} it is even conjectured that Hagedorn behavior of the thermal density of states and center symmetry are tied to each other.) When $\beta \sim R$, Hagedorn instabilities may drive a phase transition to a deconfined phase, depending on the matter content and the boundary conditions for fermions on $S^1$. The reason is that using periodic boundary conditions for fermions inserts $(-1)^F$ into the partition function, and this can result in cancellations that lead to the elimination of Hagedorn instabilities. Naively one might have thought that in non-supersymmetric systems the existence of Hagedorn scaling in the density of states would necessarily force deconfinement transitions regardless of boundary conditions, but this is not always true, as emphasized in Refs.~\cite{Basar:2013sza,Basar:2014jua}. Even in non-supersymmetric systems, there are sometimes remarkable cancellations between bosonic and fermionic states which end up preserving confinement for any $\beta$. These cancellations are associated with emergent large-$N$ fermionic symmetries and large-$N$ volume independence~\cite{Basar:2013sza}. 

On general grounds, we expect the confined phase of such large-$N$ theories to be describable as weakly-coupled string theories.  We note, however, that on $S_R^3 \times S_{\beta}^1$ the energy $E$ of states at excitation level $n$ is given by
\begin{align}
E(n) = n/R \,, 
\end{align}
while it can be shown that the thermal density of states $\rho(n)$ scales as~\cite{Sundborg:1999ue,Aharony:2003sx}
\begin{align}
\rho(n) \sim e^{+\beta_H n}  ~~ {\rm as} ~~ n\to\infty~.
\label{eq:HagedornScaling}
\end{align}
This should be contrasted with the behavior of free string theories in flat space, where $E(n) \sim \sqrt{n}$ while $\rho \sim e^{\beta_H \sqrt{n}}$.  Here, however, we are far from the flat-space limit, since the effective string tension $\sim 1/R$ that one would infer from the spectrum is of the same magnitude as the curvature of the $S^3 \times S^1$ spacetime. Consequently we find the asymptotic behavior indicated in Eq.~\eqref{eq:HagedornScaling}.

%%%%%%%%%%%%%%%%%%%%%%%%%%%%%%%%%%%%%%
\section{Modularity of large-$N$ partition functions}
\label{sec:ModularForms}
%%%%%%%%%%%%%%%%%%%%%%%%%%%%%%%%%%%%%%

In this section we show that the partition functions of adjoint-matter confining gauge theories on $S^3 \times S^1$ at large $N$ and $\lambda=0$ can be rewritten as finite products of modular forms and Jacobi forms in the variable $\tau$. The fact that this rewriting is possible is one of our central results. Since the chiral torus partition functions of 2D CFTs are finite products of modular forms, this is a key piece of evidence for the relation in Eq.~\eqref{eq:TheClaim}.  In this regard, our results here generalize those of Refs.~\cite{Basar:2014jua,Basar:2015xda}. The results of this section also have some overlap with those of Ref.~\cite{Zuo:2015mxk}, which appeared as this paper being prepared for submission.

As a warm-up, in Sect.~\ref{sec:TheIndex} we show that the $\mathcal{N}=4$ superconformal index can be written as a finite product of modular forms at large $N$. Sect.~\ref{sec:ModularExpressions} contains a demonstration that the partition functions of generic adjoint-matter theories can be written as modular forms at large $N$, while Sect.~\ref{sec:SUSY1} explains how to write confined-phase partition functions as modular forms in the exceptional case of QFTs that would be supersymmetric in the flat-space limit.  Finally, in Sect.~\ref{sec:Bosonic}, we shall see that the modular-form representation of the partition functions of theories with only bosonic matter fields simplifies in particularly significant way~\cite{Basar:2015xda}.

%%%%%%%%%%%%%%%%%%
\subsection{Large-$N$ superconformal index}
\label{sec:TheIndex}
%%%%%%%%%%%%%%%%%%
As described in Refs.~\cite{Kinney:2005ej,Romelsberger:2005eg}, the $\mathcal{N}=4$ superconformal index $\mathcal{I}$ is an $S^3 \times S^1$ partition function for $\mathcal{N}=4$ supersymmetric Yang-Mills (SYM) theory, where the theory is coupled to the curvature in such a way that some of the supercharges are unbroken. By construction, $\mathcal{I}$ is a kind of Witten index, and does not depend on the 't~Hooft coupling $\lambda$. The gauge theory has an $SO(4) \simeq SU(2)_{1} \times SU(2)_{2}$ isometry group for $S^3$, associated with two conserved Cartan angular momentum charges $j_{1,2}$; a $U(1)$ isometry group for $S^1$, associated with the energy $E$; and a global $SU(4)$ R-symmetry, associated with three conserved Cartan charges $R_{i}, i=1,2,3$. The $\mathcal{N}=4$ superconformal index $\mathcal{I}$ depends on four continuous parameters $T, V,W,Y$
as
\begin{align}
\mathcal I(T,Y,V,W) = \tr (-1)^F T^{2(E + j_1)} Y^{2j_2} V^{R_2} W^{R_3}~.
\end{align}
At large $N$, the superconformal index can be written via Eq.~(4.7) of Ref.~\cite{Kinney:2005ej}: 
\begin{eqnarray}
\mathcal I(T,Y,V,W) = \prod_{n=1}^\infty {1\over 1-f(T^n,Y^n,V^n,W^n)}
\label{eq:SuperconformalIndex}
\end{eqnarray}
where
\begin{align}
1-f(T,Y,V,W) = {(1-T^2 V)(1-T^2 W/V)(1-T^2 /W)\over (1-T^3 Y)(1-T^3/Y)}.
\end{align}
One way to derive this expression is by explicitly counting the states which can contribute to the index, with attention paid to the $U(N)$ singlet constraint. Another approach to finding $\mathcal{I}$ proceeds by evaluating a path-integral on $S^3 \times S^1$ with certain fugacities turned on, in the $\lambda \to 0$ limit. The only mode which remains massless on $S^3$ is the holonomy of the Wilson loop wrapping $S^1$. Integrating out all other (massive) modes yields a matrix model which determines an effective potential for the eigenvalues of the Wilson loop. Eq.~\eqref{eq:SuperconformalIndex} results from the observation that this one-loop effective potential is minimized by a center-symmetric eigenvalue distribution for all $\beta/R$ and evaluating the Gaussian integral around this configuration. The Gaussian approximation becomes exact at large $N$. The large-$N$ limit of $\mathcal{I}$ can be thought of as a `confining-phase' partition function, in the limited sense that it is associated with a center-symmetric holonomy for the color gauge field.

We now point out that for generic values of $T,V,W,Y$, Eq.~\eqref{eq:SuperconformalIndex} can be re-expressed in terms of objects with known modular transformations. To do this we first parametrize $V,W, Y$ as
\begin{align}
V = T^v ,\qquad W = T^{w},\qquad Y = T^y,
\end{align}
and then define the modular parameter $\tau$ via
\begin{align}
T = e^{-\beta/2R} = e^{2\pi i \tau}.
\end{align}
One can associate the imaginary part of $\tau$ with a ratio of the circumferences of $S^1$ and $S^3$: $\mathrm{Im} \, \tau = \frac{\beta}{2 \pi (2R)}$. The physical interpretation of $\re \tau$ within the index is less clear; our expression above amounts to analytically continuing $T = e^{-t}, t \in [0,1)$ to $T = e^{2\pi i \tau}, \tau \in \mathbb{H}$. With these identifications, we obtain 
\begin{align}
\mathcal I(T,Y,V,W) &=\prod_{n=1}^\infty \frac{(1-T^{(3+y) n})(1-T^{(3-y)n})}{(1-T^{(2+v) n})(1-T^{(2+w-v)n})(1-T^{(2-w) n})}\nonumber\\
&= \frac{\eta\left((3+y)\tau\right)\eta\left((3-y)\tau\right)}{\eta\left((2+v)\tau\right)\eta\left((2+w-v)\tau\right)\eta\left((2-w)\tau \right)} ,
\label{eq:IndexModularGeneral}
\end{align}
where we have used the product representation of the Dedekind $\eta$ function. The fact that such an expression is available is non-trivial, because it means that the energies and degeneracies of the states contributing to $\mathcal{I}$ are essentially those of a finite collection of two-dimensional free field theories. We note that already at finite $N$, it is known that the Schur limit of the superconformal index is controlled by a 2D chiral algebra~\cite{Beem:2013sza}, and consequently Schur limits of superconformal indices have a modular structure~\cite{Razamat:2012uv,Beem:2013sza,Cordova:2015nma,Bourdier:2015sga,Bourdier:2015wda}. It would be very interesting to understand the relation between our simple observations about the large-$N$ limit of the superconformal index of Ref.~\cite{Kinney:2005ej}, and the detailed discussions of modularity in superconformal indices in Ref.~\cite{Beem:2013sza}.

The result in Eq.~\eqref{eq:IndexModularGeneral} has several interesting and useful properties. For instance, it allows a Cardy-like~\cite{Bloete:1986qm} relation between the small-$\beta$ and large-$\beta$ behaviors of the large-$N$ limit of the index. (For an interesting discussion of Cardy-like relations for superconformal indices at finite $N$, see Ref.~\cite{DiPietro:2014bca}.) The asymptotics of $\mathcal{I}$ can be read off from the appropriate asymptotics of the $\eta$ functions, bearing in mind that the small- and large-$\beta$ asymptotics are related by modular transformations acting on the argument of each $\eta$ function. We refer to the resulting relation as ``Cardy-like" because the index is modular covariant, in the sense of being built out of modular forms, but is not modular invariant. Consequently, the relation between small- and large-$\beta$ asymptotics is more complicated than in Ref.~\cite{Bloete:1986qm}.  

First, at large $\beta$, i.e., at large $\im \tau$, we have
\begin{align}
\eta( (3+y) \tau) &\sim e^{2\pi i (3+y)\tau/24} \;, \qquad
\eta( (3-y) \tau) \sim  e^{2\pi i (3-y)\tau/24} \;, \qquad
\eta( (2+v) \tau) \sim  e^{2\pi i (2+v)\tau/24}\:, \nonumber\\
&\eta( (2+w-v) \tau) \sim  e^{2\pi i (2+w-v)\tau/24} \;, \qquad
\eta( (2-w) \tau) \sim  e^{2\pi i (2-w)\tau/24} .
\end{align}
Putting these asymptotics together, we see that at large $\beta$ (i.e., at large $\im \tau$), we have 
\begin{align} 
\lim_{\beta\to 0} \, \mathcal{I}(\beta) = 1.
\end{align}
To say this another way, each $\eta$ function has a vacuum energy which is dictated by its modular properties, and the combination of vacuum energies relevant to the index is
\begin{align} 
E_{\rm vac} = \frac{1}{24} \left[(3+y) + (3-y) -(2+v) -(2+w-v)-(2-w) \right]= 0.
\label{eq:IndexCasimir}
\end{align}
Not coincidentally, $E_{\rm vac}=0$ is also the result predicted by T-reflection symmetry~\cite{Basar:2014mha}.   We hasten to make two comments for readers who wish to compare our result to results in some of the prior literature~\cite{Assel:2014paa,Assel:2014tba,Lorenzen:2014pna,Assel:2015nca,Ardehali:2015hya,Ardehali:2014esa,Bobev:2015kza}.  It is correct to call Eq.~\eqref{eq:IndexCasimir}  the Casimir energy given two assumptions.  One is that the large-$N$ limit is taken before the removal of the UV cutoff (which must be introduced at intermediate stages in calculating vacuum energies). The other is that we assume that the renormalization scheme being used is consistent with the modular properties of the large-$N$ spectrum, as expressed in Eq.~\eqref{eq:IndexModularGeneral}. If we were to shift the Casimir vacuum energy in the large-$N$ QFT away from zero to $\Delta$, we would find that $\mathcal{I}$ could not be written directly as a combination of modular forms. In such a case, we would get a remaining factor of $q^{\Delta}$ in Eq.~\eqref{eq:SuperconformalIndex}. For a more detailed discussion of the computation of vacuum energies at large $N$ and the implications of modularity, see Sect.~\ref{sec:VacuumEnergy}.

Second, the modular properties of the $\eta$ functions imply that for small $\beta$, i.e., for small $\im \tau$, we have
\begin{align}
\eta( (3+y)\tau) &\sim e^{-\pi i /12(3+y)\tau} \;, \qquad
\eta( (3-y)\tau) \sim e^{-\pi i /12(3-y)\tau} \;, \qquad
\eta( (2+v)\tau) \sim e^{-\pi i /12(2+v)\tau} \;, \nonumber\\
&\eta( (2+w-v)\tau) \sim e^{-\pi i/12 (2+w-v)\tau} \;, \qquad
\eta( (2-w)\tau) \sim e^{-\pi i /12(2-w)\tau} .
\end{align}
This allows us to establish that for small, purely imaginary $\tau$ (equivalently, for small $\beta$), the index behaves as
\begin{eqnarray}
\lim_{\beta\to 0} \mathcal{I}(\beta) &\sim& \exp\left[\frac{i \pi \left(\frac{1}{v-w-2}-\frac{1}{v+2}+\frac{1}{w-2}+\frac{1}{3-y}+\frac{1}{y+3}\right)}{12 \tau}\right] \nonumber\\
& =& \exp\left[\frac{\pi ^2 R \left(\frac{1}{v-w-2}-\frac{1}{v+2}+\frac{1}{w-2}+\frac{1}{3-y}+\frac{1}{y+3}\right)}{3 \beta}\right]~.
\end{eqnarray}
This follows the characteristic 2D behavior summarized in Eq.~\eqref{eq:2DBehavior}, rather than the small-circle behavior one might expect from Eq.~\eqref{eq:4Dbehavior} for a generic 4D theory. In this case, the lack of a $\beta^{-3}$ divergence in $\log \mathcal{I}$ is easy to understand: it is simply due to supersymmetry~\cite{DiPietro:2014bca}. For any QFT with a $(-1)^F$-twisted partition function $\tilde{Z}$, the coefficient of $\beta^{-3}$ in $\log\tilde{Z}$ can be related to the coefficient of the quartic UV-cutoff divergence in the vacuum-energy spectral sum of the theory. But in supersymmetric field theories, this divergence is absent, and so the $\beta^{-3}$ coefficient must vanish.  It then follows that the small-$\beta$ expansion of $\log \tilde{Z}$ begins as $\beta^{-1}$. However, the reason for the vanishing of the coefficient of $\beta^{-3}$ is more subtle in our manifestly non-supersymmetric examples below.

The relation between the spectrum encoded in the large-$N$ superconformal index and the spectrum of a 2D theory can be made much sharper, at least for certain choices of fugacities. Let us consider a simple one-parameter slice through the space of fugacities, defined by setting
\begin{align}
 v = 1-y,\qquad w= (1-y)/2,
\label{eq:FugacitySlice}
\end{align}
and let us denote the resulting index as $\mathcal{I}(\tau,y)$. The small-$|\tau|$ asymptotics derived above simplify to $\mathcal{I}(\tau,y) \to \exp\left[-\frac{2\pi i }{16 \tau (y+3)/2}\right]$, and the index can now be written as
\begin{align}
\mathcal I(\tau,y) &= \frac{\eta\left( (y+3)\tau\right)}{\eta\left( \half(y+3)\tau\right)^2} = \frac{1}{\sqrt{2}}\frac{1}{\eta\left(\half (y+3)\tau\right)}\left[\frac{\Th{\half}{0}\left(\half(y+3)\tau\right)}{\eta\left(\half(y+3)\tau\right)}\right]^{\half} .
\end{align}
Introducing a modified modular parameter $\tilde{\tau} \equiv \half(y+3)\tau$, we thus see that the index takes the form
\begin{align}
\mathcal I(\tilde{\tau})&= \frac{1}{\sqrt{2}}\frac{1}{\eta\left(\tilde{\tau}\right)}
\left[\frac{\Th{\half}{0}\left(\tilde{\tau}\right)}{\eta(\tilde{\tau})}\right]^{\half}\,.
\label{eq:IndexModular}
\end{align}

We are now in a position to give our first explicit illustration of the 4D-2D relation advertised in the Introduction. First, recall that the left-moving sector of a $c=1$ non-compact free scalar CFT on a torus with modular parameter $\tilde{\tau}$ has a partition function given by $[\eta(\tilde{\tau})]^{-1}$. Second, recall that the left-moving sector of a $c=1/2$ free fermion CFT on a torus with NS-R boundary conditions has a partition function given by $\big\{\Th{\half}{0}\left(\tilde{\tau}\right)/\eta(\tilde{\tau})\big\}^{1/2}$. A direct product of these CFTs is a supersymmetric CFT.~ 
Thus, evaluating the total trace over the Hilbert space of, e.g., the left-moving degrees of freedom yields a (chiral) partition function of the 2D CFT:
\begin{align}
Z_{\rm 2D} = \frac{1}{\eta(\tilde{\tau})}\left[\frac{\Th{\half}{0}\left(\tilde{\tau}\right)}{\eta(\tilde{\tau})}\right]^{\half}.
\label{eq:2DCFTindex}
\end{align}
Comparing Eq.~\eqref{eq:2DCFTindex} to Eq.~\eqref{eq:IndexModular}, we thus find the relation
\begin{align}
\mathcal I(\tilde{\tau}) =Z_{\rm 2D}(\tilde{\tau}) ,
\end{align}
which matches the general form of Eq.~\eqref{eq:TheClaim}. Of course, our identification of a specific 2D CFT associated to $\mathcal I(\tilde{\tau})$ is not unique, since there may be many distinct QFTs with coincident spectra. It is nevertheless interesting that an identification between the partition functions of 4D and 2D CFTs is possible at all, given that 4D-2D isospectralities are not expected for the reasons already mentioned in the Introduction.   

In the case of the superconformal index, the large-$N$ equivalence between the 4D and 2D theories extends beyond the spectrum.  The reason is that derivatives of the 4D partition function with respect to the chemical potential $y$ yield correlation functions of the conserved charge which couples to $y$.  Since the modular parameter $\tilde{\tau}$ of the 2D theory has a known dependence on $y$, this allows one to relate at least some correlation functions in the 4D theory to observables of the 2D theory.  

%%%%%%%%%%%%%%%%%%%%%%%%%%%%%%%%%%%%%%%%%%%%%%%
\subsection{Confining theories with generic matter content}
\label{sec:ModularExpressions}
%%%%%%%%%%%%%%%%%%%%%%%%%%%%%%%%%%%%%%%%%%%%%%%

We now turn back to generic adjoint-matter large-$N$ QFTs on $S^3 \times S^1$, with either periodic or anti-periodic boundary conditions for fermions, and show how Eq.~\eqref{eq:TheClaim} arises in this context. More precisely, we now show that the partition functions in Eq.~\eqref{eq:Zt} can be rewritten as a finite product of modular forms and Jacobi theta-functions, with a modular parameter $\tau = t_1 + i t_2$ defined at the end of Sect.~\ref{sec:PartitionFunctions}. These results hold for any $n_f, n_s$. Crucially, the modularity properties we find are not tied to supersymmetry. Supersymmetric cases occur where $n_f = \kappa +1, n_s = 2 \kappa, \kappa \in \mathbb{N}$, which corresponds to $\mathcal{N}=1$ SYM theory with $\kappa$ adjoint matter supermultiplets. 

%%%%%%%%%%%%%%
\subsubsection{$(-1)^F$-twisted partition functions}
\label{sec:twistedPFs}
%%%%%%%%%%%%%%

Let us introduce the shorthand notation $Q \equiv q^{1/2}$, and start our analysis with $(-1)^F$-twisted partition functions. These partition functions can be written as 
\begin{align}
\tilde{Z}(\tau)&=\prod_{n=1}^\infty \frac{\left(1-Q^{2 n}\right)^3}{(1+Q^{6 n})-(n_s+3) (Q^{2 n}+ Q^{4 n})+4n_f\,Q^{3 n}}
=\prod_{n=1}^\infty \frac{\left(1-Q^{2 n}\right)^3}{\tilde{P}_{\rm twisted}(Q^n)} \, .
\label{eq:twistedPF}
\end{align}

{\it A priori}\/, if $\tilde{P}_{\rm twisted}(Q)$ were a generic sixth-order polynomial, we would have hope of being able to write Eq.~\eqref{eq:twistedPF} in terms of modular forms with closed-form expressions for their parameters. However, the six roots of the polynomial $\tilde{P}_{\rm twisted}(Q)$ turn out to be a set of three pairs of numbers which are reciprocals of each other. This is a consequence of the T-reflection symmetry discussed in Sect.~\ref{sec:PartitionFunctions}; see Appendix~\ref{sec:RootAppendix} and Ref.~\cite{Basar:2014mha} for a full discussion. Thus, $\tilde{P}_{\rm twisted}(Q)$ can be factored as 
\begin{align}
\tilde{P}_{\rm twisted}(Q) &= \prod_{\alpha=1}^3(Q+z_{\alpha})(Q+1/z_{\alpha})  \, .
\label{eq:Ptwisted}
\end{align}

With this factorization in hand, we now use the product representations of the Dedekind $\eta(\tau)$ function and the elliptic $\vartheta$ functions with characteristics $\Th{\alpha}{\beta}(\tau)$, tabulated in Appendix~\ref{sec:AppendixConventions}, to rewrite $\tilde{Z}(\tau)$ in a way that exposes its modular properties: 
\begin{align}
\tilde{Z}(\tau)
&= \prod_{n = 1}^{\infty} \prod_{\alpha = 1}^{3} \frac{ (1-Q^{2n})}{ (1 + Q^n z_{\alpha} ) (1 + Q^n/z_{\alpha} ) } 
\nonumber  \\
&= \prod_{\alpha = 1}^{3} \prod_{n = 1}^{\infty} 
\frac{ (1-Q^{2n})}{ (1 + Q^{2n} z_{\alpha} ) (1 + Q^{2n} /z_{\alpha} )} \frac{1}{ (1 + Q^{2n-1} z_{\alpha} ) (1 + Q^{2n-1}/ z_{\alpha} ) } \nonumber\\
&= \prod_{\alpha = 1}^{3} \prod_{n = 1}^{\infty} 
\frac{ (1-q^n)}{ (1 + q^n z_{\alpha} ) (1 + q^n /z_{\alpha} )} \frac{1}{ (1 + q^{n-1/2} z_{\alpha} ) (1 + q^{n-1/2}/ z_{\alpha} ) }  \nonumber \\
&= \prod_{\alpha = 1}^{3} \frac{ 2 \cos(\pi b_{\alpha})
\eta(\tau)^3 }{ \theta_{2}(b_{\alpha},\tau)}
 \frac{1}{ \theta_{3}(b_{\alpha},\tau) } 
\label{eq:multiline}
\end{align}
where $z_{\alpha} = e^{2 \pi i b_{\alpha}}$ and again $q = e^{-\beta/R} \to e^{2 \pi i \tau}$. Note that in passing between the first and second lines of Eq.~\eqref{eq:multiline}, we have split the product into a product over even and odd integers $n$. Likewise, in passing between the third and fourth lines of Eq.~\eqref{eq:multiline}, we have assumed that $z_{\alpha} \neq -1$ (or $b_\alpha\not= 1/2$). This assumption holds for generic $n_f$ and $n_s$, but fails for certain special values of $n_f$ and $n_s$. We shall discuss the cases with $z_\alpha= -1$ in Sect.~\ref{sec:SUSY1}.  

For the rest of our analysis, it will be convenient to rewrite this result as
\begin{align}
\tilde{Z}(\tau) = \prod_{\alpha = 1}^{3} \left[ 2 e^{-i \pi b_{\alpha}}\cos(\pi b_{\alpha}) \, \eta(\tau)^2 \frac{1}{\eta(\tau)}
\frac{ \eta(\tau) }{ \Th{1/2}{b_{\alpha}}(\tau)}
\frac{\eta(\tau)}{\Th{0}{b_{\alpha}}(\tau) } \right] \,,  \label{twistZ0} 
\end{align}
where have again used the assumption $z_{\alpha} \neq -1$. The expression in Eq.~(\ref{twistZ0}) is one of our key results.  As we see from Eq.~(\ref{twistZ0}), this expression is a finite product of modular forms and Jacobi forms. Consequently, this establishes one of our main claims: $\tilde{Z}(\tau)$ is a (component of a vector-valued, meromorphic) modular form at $N =\infty$, with modular weight $+3/2$. For more on the modular properties of $\tilde{Z}(\tau)$, see Sect.~\ref{sec:ModularCompletion}.

%%%%%%%%%%%%
\subsubsection{Thermal partition functions}
\label{sec:thermalPFs}
%%%%%%%%%%%%

We now turn to the thermal partition functions. The infinite-product representation of the thermal partition function can be obtained from Eq.~\eqref{eq:twistedPF} by using $Z(Q) = \tilde{Z}(-Q)$. The presence of a factor of $(-1)^n$ in front of the fermion terms in the infinite products in thermal partition functions introduces a minor subtlety for rewriting the infinite product using modular forms. To illustrate this, we observe that 
\begin{align}
Z(\tau)&=\prod_{n=1}^\infty \frac{\left(1-Q^{2 n}\right)^3}{(1+Q^{6 n})-(n_s+3) (Q^{2 n}+ Q^{4 n})+(-1)^n 4n_f\,Q^{3 n}}
\nonumber\\
&=\prod_{n=1}^\infty \frac{\left(1-Q^{2 n}\right)^3}{\tilde{P}_{\rm twisted}(Q^{2n})}\frac{1}{P_{\rm thermal}(Q^{2n-1})} \, .
\label{eq:thermalPF}
\end{align}
This makes it clear that for even $n$, the analytic structure is controlled by the polynomial $\tilde{P}_{\rm twisted}$ we saw before, while for odd $n$, the analytic structure is controlled by 
\begin{align}
P_{\rm thermal}(\tau) &= \prod_{\alpha=1}^3(Q-z_{\alpha})(Q-1/z_{\alpha})  \, .
\label{eq:Pthermal}
\end{align}
We are now in a position to rewrite Eq.~\eqref{eq:thermalPF} in terms of modular forms. We obtain
\begin{align}
Z(\tau)
&= \prod_{n = 1}^{\infty} \prod_{\alpha = 1}^{3}
\frac{ (1-Q^{2n})}{ (1 + Q^{2n} z_{\alpha} ) (1 + Q^{2n}/z_{\alpha} ) } 
\frac{ 1}{ (1 - Q^{2n-1} z_{\alpha} ) (1 - Q^{2n}/z_{\alpha} ) } \nonumber\\
&= \prod_{n = 1}^{\infty} 
\prod_{\alpha = 1}^{3} \frac{ (1-q^n)}{ (1 + q^n z_{\alpha} ) (1 + q^n /z_{\alpha} )}
\frac{1}{ (1 - q^{n-1/2} z_{\alpha} ) (1 - q^{n-1/2}/ z_{\alpha} ) } \label{eq:modStep3} \nonumber\\
&= \prod_{\alpha = 1}^{3} \frac{ 2 \cos(\pi b_{\alpha})\eta(\beta/R)^3 }{ \theta_{2}(b_{\alpha},\beta/R)} 
\frac{1}{\theta_{4}(b_{\alpha},\beta/R) } ,
\end{align}
which we rewrite as
\begin{align}
Z(\tau) = \prod_{\alpha = 1}^{3} \left[ 2 e^{-i \pi b_{\alpha}}\cos(\pi b_{\alpha}) \, \eta(\tau)^2 \frac{1}{\eta(\tau)}
\frac{ \eta(\tau) }{ \Th{1/2}{b_{\alpha}}(\tau)}
\frac{\eta(\tau)}{\Th{0}{b_{\alpha}+\half}(\tau) } \right] . \,  \label{thermZ0} 
\end{align}
Once again, in obtaining these results we have assumed that $z_\alpha\not = -1$ (i.e., $b_\alpha \not= 1/2$).

Like the expression in Eq.~(\ref{twistZ0}), this expression is another of our key results and has well-defined behavior under modular transformations. We thus conclude that the confined-phase large-$N$ partition functions of generic 4D adjoint-matter gauge theories on $S^3 \times S^1$ in the $\lambda \to 0$ limit are (components of vector-valued) modular forms, with modular weight $+3/2$.  Furthermore, consulting the conventions laid out in Appendix~\ref{sec:AppendixConventions}, we see that the modular $T: \tau \to \tau+1$ transformation exchanges the functions which distinguish $\tilde{Z}(\tau)$ from $Z(\tau)$, i.e., $T: \Th{0}{b_{\alpha}+1/2}(\tau) \leftrightarrow \Th{0}{b_{\alpha}}(\tau)$.  Thus, the modular $T$-translation maps the twisted and thermal partition functions to each other:
\begin{equation}
T : ~~~  Z ~ \longleftrightarrow \tilde Z~.
\end{equation}
More details concerning the behavior of $Z(\tau)$ and $\tilde{Z}(\tau)$ under modular transformations are discussed in Sect.~\ref{sec:ModularCompletion}.

%%%%%%%%%%%%%%%%%%%%%%%%%%%%%%%%%%%%%
\subsection{Confining theories with supersymmetric matter content}
\label{sec:SUSY1}
%%%%%%%%%%%%%%%%%%%%%%%%%%%%%%%%%%%%%

While for generic choices of $n_f$ and $n_s$ the modular weight of the large-$N$ partition functions is $+3/2$, it is different for theories with $\mathcal{N}=1$ supersymmetry. In such theories, $n_f$ and $n_s$ are related by $n_f = \kappa+1$ and $n_s = 2\kappa$, where $\kappa \ge 0$ is the number of adjoint $\mathcal{N}=1$ matter multiplets.  

It is easy to see why these cases require special treatment. When $n_f = \kappa+1$ and $n_s = 2\kappa$ with $\kappa \ge 1$, the twisted polynomial, which dictates the pole structures common to both the twisted and thermal partition functions at even $n$ in Eqs.~\eqref{eq:twistedPF} and \eqref{eq:thermalPF}, simplifies to
\begin{align}
\tilde{P}^{\kappa}_{\rm twisted}(\tau) &= (Q+(-1))^2 \left[ (Q^4+1) + 2(Q+Q^3) - 2 \kappa Q^2 \right].
\label{eq:SUSYmod1}
\end{align}
So for $\kappa = 0, 1, 2$ we see that $\tilde{P}^{\kappa}_{\rm twisted}(Q)$ has a second-order root at $Q = 1$. In the notation of the previous section, this implies that one of roots takes the value $z_{\alpha} = -1$ due to the $(1-Q)^2$ factor present for typical supersymmetric theories. For the even more special case of $\kappa = 3$, corresponding to $\mathcal{N}=4$ SYM, the root at $Q=1$ becomes fourth-order.

The position of the root at $z_\alpha= -1$, corresponding to $\beta_\alpha = 1/2$, is the source of the difficulty. As we have seen, all of our partition functions contain the expression
\begin{equation}
\frac{ \cos(\pi b_{\alpha})}{ \Th{\half}{b_{\alpha}}(\tau)}~. 
\label{theproblem}
\end{equation}
However, while this expression for $b_\alpha\not=1/2$ is perfectly reasonable and straightforward to interpret, for $b_\alpha= 1/2$ we find that both the numerator and the denominator vanish identically. Indeed, with $b_\alpha= 1/2$ the denominator becomes nothing but $\Th{\half}{\half}(\tau) = \theta_1(0,\tau)=0$. Thus, for $z_\alpha= -1$ or $\beta_\alpha=1/2$, our previous expressions become indeterminate.

There are two ways to proceed, which give the same result.  One way is to look directly at the infinite-product expressions for the supersymmetric cases and read off their expressions in terms of modular forms and Jacobi forms.  The second way is to obtain modular expressions for these special cases by taking a limit of the modular expressions valid for generic $n_f$ and $n_s$.  Due to the subtlety highlighted above, we do this by identifying
\begin{equation}
\frac{ \cos(\pi/2)  }{ \Th{\half}{\half}(\tau)}  = \lim_{b_\alpha\to 1/2} ~\frac{ \cos(\pi b_{\alpha})}{ \Th{\half}{b_{\alpha}}(\tau)} = \lim_{b_\alpha\to 1/2} ~\frac{ \cos(\pi b_{\alpha})}{ \theta_1 (b_\alpha-1, \tau)} = {1\over 2\eta(\tau)^3}~ 
\label{Dm10}
\end{equation}
where in the final equality we have used l'H\^opital's rule along with the identity 
\begin{align}
\partial_{c} \theta_1(c,\tau)\bigg|_{c=0} &= 2 \eta(\tau)^3 \,. \label{Dm9}
\end{align}
However, we see that this final expression has modular weight $k=-3/2$, as opposed to the modular weight $k= -1/2$ of the expression in Eq.~(\ref{theproblem}) with which we started. Thus, we see that the modular weight drops by $1$ when $b_{\alpha}$ hits $\half$.  More succinctly, we have
\begin{align} 
\prod_{n = 1}^{\infty} \frac{1}{ (1 + q^n z_{\alpha} ) (1 + q^n /z_{\alpha} )} &=
\begin{cases}
\frac{ 2 \cos(\pi b_{\alpha}) q^{1/12 }\eta(\tau) }{ \Th{\half}{b}(\tau)} & z_{\alpha} \neq -1 \\
\frac{ q^{1/12} }{ \eta(\tau)^2} & z_{\alpha} = -1 
\end{cases} \, 
\label{eq:SUSYCases}
\end{align}
Thus modular-form representations for the infinite products of gauge-theory partition functions with (exceptional) roots $z_{\alpha}=-1$ have modular weights which are one unit lower than those with generic roots $z_{\alpha} \neq -1$.   

We emphasize that the second approach described above rests on obtaining the result for the special case $b_\alpha= 1/2$ via the formal limit $b_\alpha\to 1/2$. While this seems mathematically reasonable, we note that arbitrary real (or complex) values of $b_\alpha$ do not generally correspond to physically realizable systems, because this amounts to allowing $n_f$ and $n_s$ to be non-integral.
  
Combining these observations, we see that the modular weight of our overall expression drops by $1$ whenever a pair of roots of $\tilde{P}^{\kappa}_{\rm twisted}(Q)$ hits $Q=1$. It can be shown that this singular locus in parameter space is given by the line $n_s = 2n_f-2$, corresponding to theories with supersymmetric matter content. Except at $n_s=6$, there is a single pair of roots at $Q=1$ along this line. Exactly at $n_s = 6$ --- corresponding to the matter content of $\mathcal{N}=4$ SYM theory --- there are two pairs of roots at $Q = 1$.  Thus the modular weight of the partition functions of theories with supersymmetric matter content is $1/2$ rather than $3/2$ for $\kappa = 0, 1, 2$. 

For $\kappa=3$, we see that $P_{\rm twisted}$ has a \emph{quadruple} root corresponding to $z_1 = z_2 = -1$, which triggers a further reduction\footnote{One may wonder if there is an even more special theory which has $\tilde{P}_{\rm twisted}(Q) = (1-Q)^6$. This does not seem possible in the set of theories we consider. Expanding out this putative polynomial yields $\tilde{P}_{\rm twisted}(Q) = (1+Q^6) + 15(Q^2+Q^4) - 20 Q^3 - 6(Q+Q^5)$. The term $(Q+Q^5)$ cannot arise for any $n_f, n_s$, even if we allow $n_s, n_f$ to be arbitrary complex numbers.  We also note that if one were to find a theory with $\tilde{P}_{\rm twisted}(q) = (1-q)^6$, the resulting partition function could be written using Dedekind $\eta$ functions and $\vartheta$ functions with \emph{rational} characteristics, indicating that this would be a partition function without Hagedorn singularties for any choice of boundary conditions. See Ref.~\cite{Basar:2014jua} for details on the connection between Hagedorn growth and values of $z_{\alpha}$ with $|z_\alpha| \neq 1$.} of the modular weight of the partition function, to $-1/2$. As a result of these observations, the twisted partition functions of gauge theories with $\kappa = 0,1,2$ or $\kappa = 3$ adjoint $\mathcal{N}=1$ matter superfields on $S^3 \times S^1$ take the form\footnote{In these expressions, $b_{\pm}=b_{\pm}(\kappa)$ for $\kappa<3$ are given by $b_{\pm}(\kappa) = \frac{1}{2\pi} \cos^{-1}\left(\frac{1\pm\sqrt{1+\kappa}}{2}\right)$. For $\kappa=3$, we have $b_{\kappa=3} = b_{+}(3) = \frac{1}{2\pi} \cos^{-1}(2)$; by contrast, $b_{-}(3) = \half$ and thus Eq.~\eqref{eq:SUSYCases} must be used in order to derive the given modular-form representation. The characteristic-dependent prefactors are simply given by the product $\prod_{\alpha} 2\cos(\pi b_{\alpha})e^{-i \pi b_{\alpha}}$. These numerical factors are algebraic numbers for all $n_s, n_f \in \{0,1,2, \ldots \}$.} 
\begin{align}
\tilde{Z}_{\kappa < 3}(\tau) &= \eta(\tau) 
\left( \frac{\eta(\tau)}{ \Th{0}{\frac{1}{2}}(\tau)} \right)
\prod_{\pm} \frac{ 2\cos(\pi b_{\pm}) e^{-i \pi b_{\pm}} \,\, \eta(\tau)^2}{ \Th{1/2}{b_{\pm}}(\tau) \,\, \Th{0}{b_{\pm}}(\tau)} \,  
\nonumber\\
\tilde{Z}_{\kappa = 3}(\tau) &=
\frac{1}{\eta(\tau)}\left(\frac{\eta(\tau)}{ \Th{0}{\frac{1}{2}}(\tau)}\right)^2
\frac{ 2\cos(\pi b_{\kappa=3}) e^{- i \pi b_{\kappa=3} }\,\, \eta(\tau)^2}{ \Th{1/2}{b_{\kappa=3}}(\tau) \,\, \Th{0}{b_{\kappa=3}}(\tau)} 
\label{eq:ZtmaxSYM} \, ,
\end{align}
while the thermal partition functions are
\begin{align}
Z_{\kappa < 3}(\tau) &= \eta(\tau)
\left( \frac{\eta(\tau)}{ \Th{0}{0}(\tau)} \right)
\prod_{\pm} \frac{ 2\cos(\pi b_{\pm}) e^{-i \pi b_{\pm}} \,\, \eta(\tau)^2}{ \Th{1/2}{b_{\pm}}(\tau) \,\, \Th{0}{b_{\pm}+\frac{1}{2}}(\tau)} \,  
\nonumber\\
Z_{\kappa = 3}(\tau) &=\frac{1}{\eta(\tau)}
\left(\frac{\eta(\tau)}{ \Th{0}{0}(\tau)}\right)^2
\frac{ 2\cos(\pi b_{\kappa=3}) e^{- i \pi b_{\kappa=3} }\,\, \eta(\tau)^2}{ \Th{1/2}{b_{\kappa=3}}(\tau) \,\, \Th{0}{b_{\kappa=3}+\frac{1}{2}}(\tau)} 
\label{eq:ZmaxSYM} \, .
\end{align}

In summary, then, the confined-phase partition functions of large-$N$ gauge theories with $0 \leq \kappa < 3$ adjoint matter supermultiplets have modular weight $+1/2$. The theory with $\kappa=3$ adjoint matter multiplets, $\mathcal{N}=4$ SYM theory, has modular weight $-1/2$.

%%%%%%%%%%%%%%%%%%%%%%%%%%
\subsection{Confining theories with purely bosonic matter}
\label{sec:Bosonic}
%%%%%%%%%%%%%%%%%%%%%%%%%%

The formulas derived in Sect.~\ref{sec:ModularExpressions} continue to apply for purely bosonic theories, with arbitrary $n_s$ and $n_f = 0$. However, for our purposes it is useful to derive shorter equivalent expressions for purely bosonic matter content. A demonstration that the expressions derived in this section are consistent with those in Sect.~\ref{sec:ModularExpressions} is given in Appendix~\ref{sec:BosonicSimplifcations}. 

The partition functions of bosonic confining large-$N$ theories can be written as
\begin{align}
Z_{\rm bosonic}(\tau) = \prod_{n=1}^{\infty} \frac{(1-q^n)^3}{(1+q^{3n})-(3+n_s)(q^n + q^{2n})} = \prod_{n=1}^{\infty} \frac{(1-q^n)^3}{ P(q^n)} \, .
\end{align}
The polynomial $P(q) = (1+q^3)-(3+n_s)(q + q^2)$ has a root at $q = -1$, and factorizes as
\begin{align}
P(q) &= (1+q)(q^2-(4+n_s)q-1) = (q+1) (q - z_b ) (q - 1/z_b ) \label{bosP}
\end{align}
where
\begin{align}
z_b &= \frac{4+n_s}{2} + \sqrt{\left( \frac{4+n_s}{2} \right)^2-1} \label{bosZ} \, .
\end{align}
Using this, we see that we can rewrite the partition function for purely bosonic theories as
\begin{align}
Z_{\rm bosonic}(\tau; n_s) = 2 \sqrt{2} ~ i \sin\left(\pi b_{n_s}\right) e^{-i\pi b} \eta(\tau)^3 \frac{\eta(\tau)}{\Th{1/2}{b_{n_s}+1/2}(\tau)} \left[\frac{\eta(\tau)}{\Th{1/2}{0}(\tau)}\right]^{1/2} \label{eq:BosonicZ}
\end{align}
where $b_{n_s} \equiv \frac{1}{2\pi}\cos^{-1}\left(2+n_s/2\right)$. Of course, Eq.~\eqref{eq:BosonicZ} still has modular weight $+3/2$, just as for the general cases represented in Eqs.~\eqref{twistZ0} and~\eqref{thermZ0}. This shows that the confined-phase large-$N$ partition functions of purely bosonic theories have the simplest structure of all of our non-trivial examples.

%%%%%%%%%%%%%%%%%%%%%%%%%%%%%%%%%%%%%%%%%%%%%%
\section{Implications of modularity of large-$N$ partition functions}
\label{sec:Implications}
%%%%%%%%%%%%%%%%%%%%%%%%%%%%%%%%%%%%%%%%%%%%%%
We have seen that the confined-phase large-$N$ partition functions of adjoint-matter gauge theories on $S^3 \times S^1$, in the $\lambda \to 0$ limit, can always be written as finite products of Dedekind $\eta$ functions and Jacobi $\vartheta$ functions. The generalization of these observations from purely bosonic Yang-Mills theory in Ref.~\cite{Basar:2015xda} to gauge theories with arbitrary numbers of adjoint scalars and adjoint fermions has several dramatic consequences.

%%%%%%%%%%%
\subsection{Vanishing vacuum energy and large-$|\tau|$ behavior}
\label{sec:VacuumEnergy}
%%%%%%%%%%%%%%%%%%%%%%

Our results imply that the large-$N$ theories we consider have vanishing vacuum energy in a renormalization scheme consistent with the symmetries of the large-$N$ spectrum, as first found by other means in  Ref.~\cite{Basar:2014hda}. 

The value of $E_{\rm vac}$ is defined as a regularized and renormalized sum over the spectrum:
\begin{align}
E_{\rm vac}=  \frac{1}{2} \sum_{n=0}^{\infty} d_n \omega_n \bigg|_{\mu_{\rm uv}} + E_{\rm{counter-terms}}(\mu_{\rm uv})
\end{align}
where $|_{\mu_{\rm uv}}$ refers to a regularization of the sum involving some high-energy scale $\mu_{\rm uv}$, and where $E_{\rm{counter-terms}}(\mu_{\rm uv})$ represents the renormalization-scheme-dependent contributions of divergent and finite counter-terms. Given a fixed regularization and renormalization scheme, $E_{\rm vac}$ is trivially well defined, but $E_{\rm vac}$ becomes most interesting if it can be shown that its value is the same for any renormalization-scheme choice consistent with the symmetries of the theory. If this happens, then $E_{\rm vac}$ will have a physical interpretation in the limit $\mu_{\rm uv} \to \infty$. We emphasize that in deciding whether an observable is scheme-dependent or not, it is vital to have a complete understanding of the symmetries of the QFT because this affects the allowed choices of scheme.  So until the constraints of possible emergent symmetries of large-$N$ confining theories are understood, it can be somewhat premature to decide whether a given quantity is scheme-dependent.

In generic 4D Poincare-invariant QFTs in finite volume, computations of $E_{\rm vac}$ using, e.g., spectral heat-kernel regulators produce a $\mu_{\rm uv}^4$ divergence. Canceling this divergence requires the introduction of a `cosmological constant' counterterm
\begin{align}
\mu_{\rm uv}^4\int d^{4} x\, \sqrt{g}\,.
\end{align} 
where $\mu_{\rm uv}$ is the UV scale. If the 4D QFT is formulated in curved space-time, one also expects a $\mu_{\rm uv}^2$ divergence related to the curvature;  this requires the addition of an `Einstein-Hilbert' counter-term
\begin{align}
\mu_{\rm uv}^2\int d^{4}x\, \sqrt{g}\, \mathcal{R}\, .
\end{align}

Without demanding scale-invariance, finite cosmological-constant terms and finite Einstein-Hilbert terms are allowed.   This means that the value of $E_{\rm vac}$ is regularization-scheme-dependent in generic non-scale-invariant 4D theories.

Our considerations focus on non-Abelian gauge theories in the free limit, $\lambda \to 0$, which are scale-invariant. Scale-invariant QFTs can only have UV divergences in $E_{\rm vac}$ which are power laws in $\mu_{\rm uv}$, which can be cancelled by the cosmological constant and Einstein-Hilbert counter-terms. Finite cosmological constant terms and finite Einstein-Hilbert terms are ruled out by scale invariance. But there is also a dimensionless term one can write when putting a theory on a curved manifold,
 \begin{align}
 b\int d^{4}x \sqrt{g} \, \mathcal{R}^2.
 \label{eq:bTermVacuum}
 \end{align} 
As emphasized in  Ref.~\cite{Assel:2015nca}, changes of $b$ produce additive shifts in the $S^3$ Casimir energy, in the same way that changing $\Lambda^4$ in $\int d^{4} x\, \sqrt{g} \Lambda^4$ produces additive shifts of the vacuum energy of non-conformal theories. All values of $b$ are consistent with 4D scale invariance. This means that in generic 4D scale-invariant theories, the value of $E_{\rm vac}$ will depend on the choice of regularization scheme. Thus, as argued in  Ref.~\cite{Assel:2015nca}, $E_{\rm vac}$ is \emph{not} a universal observable in the renormalization-group sense even in systems with scale or conformal invariance in 4D.~ It depends on the choice of renormalization scheme, related to a choice of $b$. To make $E_{\rm vac}$ a continuum observable, one needs to consider a special subclass of theories which have extra symmetries which constrain the possible values of $b$.  A prominent example of such theories are superconformal quantum field theories, as emphasized in  Ref.~\cite{Assel:2015nca}.

We now observe that $E_{\rm vac}$ appears to be a scheme-independent observable in large-$N$ gauge theories in the limit considered in this paper. The basic point is that large-$N$ 4D gauge theories in the $\lambda \to 0$ limit are non-generic 4D theories. In the limit $\lambda \to 0$, they are clearly scale-invariant, which forbids most finite counter-terms, but in principle leaves $b$ from Eq.~\eqref{eq:bTermVacuum} unfixed. The far less trivial point is that these theories have rich emergent symmetries at large $N$, as revealed by the modular structure of their partition functions. The modularity of the partition functions is consistent with only one choice of $b$, which is $b=0$. 

To show how modularity fixes the value of $b$, we first recall why the normalizations of modular forms are fixed by their modular properties. Modular forms $f(\tau)$ have $q$-series representations, $f(\tau) = q^{\Delta}\sum_{n\ge0} c_n q^n$, and one can think of $q = e^{2\pi i \tau}$ as a Boltzmann factor. Then the powers of $q$ are the energies (in natural units) of a collection of states which are related by conformal symmetry to a `primary' state with energy $\Delta$, and $f(\tau)$ is a type of chiral partition function. The individual Boltzmann factors $q^{n} = e^{2\pi i n \tau}$ are not well-behaved under the $S$-transformation, so the modular properties of $f(\tau)$ are properties of the analytic continuation of the $q$-series, rather than properties of the individual terms in the $q$-series. This means that one cannot change the coefficients $c_n$ without destroying the modular transformation properties of $f(\tau)$. It also implies that the vacuum energy $\Delta$ appearing in the definition of $f(\tau)$ cannot be be shifted. To see this, observe that if one were to shift $\Delta \to \Delta + \Delta'$, one would obtain $f(\tau)\to f'(\tau) \equiv q^{\Delta'} f(\tau)$. But $f'(\tau)$ is not a modular form unless $\Delta' = 0$ because $q^{\Delta'}$ is not a modular form unless $\Delta' = 0$. Indeed, if a function $f(\tau)$ is a modular form, its overall `vacuum energy' $\Delta$ is fixed by the modular properties and can be determined via a sum rule on $c_n$.

These observations imply that our rewriting of 4D partition functions in terms of modular and Jacobi forms is possible only for a special value of the vacuum energy $E_{\rm vac}$ of the 4D QFT --- a value which is determined by the modular transformation properties of the modular forms comprising $Z_{\rm 4D}$. These modular transformation properties are, in turn, determined by the spectrum of the theory. Given the modular properties of the spectrum, the value of $E_{\rm vac}$, calculated in regularization schemes consistent with the spectral symmetries, is thus uniquely determined.

Thus, remarkably, if one takes the constraints from the modular symmetries seriously, the value of $E_{\rm vac}$ 
for 4D large-$N$ QFTs in the $\lambda \to 0$ limit becomes a scheme-independent observable in the renormalization-group sense.  Moreover, there are in fact two more surprises.  First, we find that the value of $E_{\rm vac}$ turns out to be \emph{numerically} universal across all of our examples, meaning that $E_{\rm vac}$  is independent of $n_s$ and $n_f$. Second, and even more surprisingly, this universal result for $E_{\rm vac}$ of the large-$N$ confining gauge theories is \emph{zero}:
\begin{align}
E_{\rm vac} = 0 = 
\begin{cases}
3 \times \frac{1}{24} +\left(\frac{1}{24} - \frac{1}{8}\right) +\half \left(\frac{1}{24} - \frac{1}{8} \right),
 & \textrm{pure YM}\\
-1 \times \frac{1}{24} +2 \left(\frac{1}{24} - 0 \right) +\left(\frac{1}{24} - \frac{1}{8} \right)+ \left(\frac{1}{24} - 0 \right),    & \textrm{$\mathcal{N}=4$ SYM}\\
3 \times \left[ \frac{1}{24} +\left(\frac{1}{24} - \frac{1}{8}\right) +\left(\frac{1}{24} - 0\right)\right], & \textrm{QCD(Adj) with $N_f=2$} \\
\ldots & \ldots
\end{cases}
\label{eq:vanishingE}
\end{align}
This matches what was found in  Refs.~\cite{Basar:2014mha,Basar:2014hda} by a direct evaluation of the spectral sums involved in $E_{\rm vac}$. In view of the considerations above, our results imply that in the renormalization-scheme choice consistent with the symmetries of the large-$N$ spectrum, one must set the coefficient $b$ of $\int d^{4}x \sqrt{g} \mathcal{R}^2$ to zero. The multi-faceted universality of these results cries out for a first-principles explanation, which we hope will become understood in future work.

Before moving on, we comment on the features of the large-$N$ limit important for our result. Our interest in general is in asymptotically free theories, and as explained in Sect.~\ref{sec:LargeN} this motivates us to take the large-$N$ limit before all other limits, including the continuum limit $\mu_{\rm uv} \to \infty$. Our result for $E_{\rm vac}$ is valid with this ordering of limits. Other calculations of $E_{\rm vac}$ in large-$N$ gauge theories on $S^3 \times S^1$, both directly in field theory, as in Ref.~\cite{Aharony:2003sx}, and using gauge-gravity duality, as in  Ref.~\cite{Balasubramanian:1999re}, use a different order of limits in which $\mu_{\rm uv}$ is taken to infinity before $N$ is taken to infinity.  This procedure leads to a different result in which $E_{\rm vac} \sim N^2 \neq 0$.  This simply implies that the vacuum energy is sensitive to the ordering of limits.

%%%%%%%%%%%%%%
\subsection{Small-$|\tau|$ behavior}
\label{sec:smallTau}
%%%%%%%%%%%%%%

At high temperature, the partition functions of generic 4D QFTs on $S^3 \times S^1_{\beta}$ behave as $\mathrm{Vol}^{-1}_{\rm S^3} \log Z \sim \beta^{-3}$. The argument for this comes down to a combination of dimensional analysis and Wilsonian renormalization-group reasoning. Generic UV-complete QFTs can be thought of as describing a renormalization-group flow between a UV fixed point and an IR fixed point, both of which are scale invariant.  At very high temperature, defined as making the $S^1$ circumference $\beta$ much smaller than any other physical scale, the physics becomes well described by the UV fixed point. At the UV fixed point, $\beta$ is the only dimensionful parameter, and dimensional analysis and the extensivity of the free energy imply that $\lim_{\beta\to 0} \log Z(\beta ) \sim -\sigma \mathrm{Vol}_{\rm S^3} \beta^{-3}$ for some numerical constant $\sigma$ determined by the details of the UV fixed point. 

This prediction that $\lim_{\beta\to 0} \log Z(\beta ) \sim -\sigma \mathrm{Vol}_{\rm S^3} \beta^{-3}$ would fail if $\sigma$ were exactly zero, since then $\log Z$ would become dominated by a subleading term in its small-$\beta$ expansion. But a vanishing $\sigma$ coefficient is extremely non-generic, and can be interpreted as a loud signal for the existence of a symmetry. For instance, in Ref.~\cite{DiPietro:2014bca} it was noted that $\sigma$ vanishes if one puts supersymmetric theories on $S^3\times S^1$ in a way that preserves some supersymmetry. This is essentially because the value of $\sigma$ is related to the value of the vacuum energy of such theories in flat space, and the flat-space vacuum energy vanishes in supersymmetric theories.

If a theory on $S^3 \times S^1$ does not enjoy supersymmetry, however, one would not generally expect to find $\sigma=0$. This can be illustrated by working out the high-temperature behavior of a free scalar field on $S^3_R\times S^1_{\beta}$. Here the partition function is $Z_{\textrm{free scalar}}(\tau) = q^{1/240} \prod_{n \ge 0} (1-q^n)^{-n^2}$, where $q = e^{2\pi i \tau} = e^{-\beta/R}$. One can then verify that
\begin{align}
\lim_{\arg \tau \to \frac{\pi}{2}}\, \lim_{|\tau| \to 0} ~\log Z_{\textrm{free scalar}}(\tau) = \lim_{\arg \beta \to 0}\left[\lim_{|\beta| \to 0} \log Z_{\textrm{free scalar}}(\tau)\right] \sim  -\mathrm{Vol}_{\rm S_3} 
\frac{\pi^2}{90 \beta^3} ,\,
\end{align} 
so that $\sigma|_{\textrm{free scalar}} = \pi^2/90 \neq 0$. 

The behavior of large-$N$ confining gauge theories on $S^3 \times S^1$ turns out to be very different than that seen in generic 4D QFTs. As a consequence of the fact that our 4D confining large-$N$ partition functions $Z_{\rm 4D}$ can be rewritten as finite products of modular forms and Jacobi forms, we find that
\begin{align}
\lim_{\arg \tau \to \frac{\pi}{2}}\left[\lim_{|\tau| \to 0} \log Z_{\rm 4D}(\tau)\right] \sim - \sigma \, \frac{2\pi R}{\beta} \, ,
\label{eq:2DbehaviorDetailed}
\end{align}
for a non-vanishing constant $\sigma$ that depends on $n_s$ and $n_f$. The behavior in Eq.~\eqref{eq:2DbehaviorDetailed} looks like what we would expect in a 2D quantum field theory, rather than a generic 4D theory. 

Given these results, we now describe the derivation of the scaling rule in Eq.~\eqref{eq:2DbehaviorDetailed} in two representative examples: $\mathcal{N}=4$ SYM with periodic boundary conditions on $S^3 \times S^1$, and pure YM theory on $S^3 \times S^1$. We begin with two preliminary comments. First, since Yang-Mills theory is obviously not supersymmetric, it is not possible to view Eq.~\eqref{eq:2DbehaviorDetailed} as a consequence of supersymmetry. Second, we note that for $\mathcal{N}=4$ SYM with periodic boundary conditions, the confining partition function has no singularities when $|\tau| \to 0$ along the imaginary axis.  Thus, for the first example of twisted ${\cal N} = 4$ SYM, the two limits in Eq.~\eqref{eq:2DbehaviorDetailed} commute. However, the limits do \emph{not} commute in pure Yang-Mills theory, as this theory has Hagedorn instabilities on $\arg \tau = \pi/2$ for $|\tau| \ll 1$.  Thus Eq.~\eqref{eq:2DbehaviorDetailed} is valid only with the ordering of limits indicated. If we were to reverse the order of the limits, the small-$\beta$ physics could not be explored from within the confining phase, and we would have to work in the deconfined phase, where we would of course obtain $\lim_{|\tau| \to 0}\left[\lim_{\arg \tau\, \to\, \pi/2} \log Z^{\rm deconfined}_{\rm 4D}(\tau)\right] \sim \beta^{-3} N^2$. 

We begin our derivation of Eq.~\eqref{eq:2DbehaviorDetailed} by considering the case of $\mathcal{N}=4$ SYM with periodic boundary conditions for fermions, which has the large-$N$ partition function given in Eq.~\eqref{eq:ZtmaxSYM}. To make the notation more transparent, we relabel ${\kappa=3} \to {\mathcal{N}=4}$ below. Our task reduces to extracting the small-$|\tau|$ behavior of $\eta(\tau), \Th{1/2}{b_{\mathcal{N}=4}}(\tau), \Th{0}{b_{\mathcal{N}=4}}(\tau),\Th{0}{1/2}(\tau)$. This can be done by exploiting the behavior of these functions under the modular $S$-transformation $\tau \to -1/\tau$ and consulting the product representations given in Appendix~\ref{sec:AppendixConventions}. For $\eta(\tau)$, the fact that $\lim_{|\tau|\to\infty} \eta(\tau) \sim\exp(2\pi i \tau \times 1/24)$ and $\eta(-1/\tau) =\sqrt{-i \tau} \eta(\tau)$ implies that at small $|\tau|$ the behavior is
\begin{align}
\lim_{|\tau|\to 0} \eta(\tau) \sim (-i\tau)^{-1/2} e^{- \frac{2\pi i}{24\tau}}.
\end{align}
For $\Th{1/2}{b_{\mathcal{N}=4}}(\tau)$ the $S$ relation is $\Th{1/2}{b_{\mathcal{N}=4}}(-1/\tau) =\sqrt{-i \tau} \,e^{\pi i b_{\mathcal{N}=4}} \Th{b_{\mathcal{N}=4}}{-1/2}(\tau)$. The fact that 
$\lim_{|\tau|\to \infty} \Th{b_{\mathcal{N}=4}}{-1/2}\sim \exp(2\pi i \tau \times b_{\mathcal{N}=4}^2/2)$ then implies that
\begin{align}
\lim_{|\tau|\to 0} \Th{1/2}{b_{\mathcal{N}=4}}(\tau) \sim (-i\tau)^{-1/2} e^{-\pi i b_{\mathcal{N}=4}} e^{- \pi i b_{\mathcal{N}=4}^2/\tau}.
\end{align}
The same line of reasoning shows that 
\begin{align}
\lim_{|\tau|\to 0} \Th{0}{b_{\mathcal{N}=4}}(\tau) \sim (-i\tau)^{-1/2} e^{-\pi i b_{\mathcal{N}=4}^2/\tau}, \qquad 
\lim_{|\tau|\to 0} \Th{0}{1/2}(\tau) \sim 2(-i\tau)^{-1/2} e^{-\frac{2\pi i}{8\tau}}.
\end{align}
Putting all this together, we find that
\begin{align}
\lim_{|\tau|\to 0}
\tilde{Z}_{\mathcal{N} = 4}(\tau) \sim \half \cos(\pi b_{\mathcal{N}=4}) (-i\tau)^{1/2} e^{\frac{ i \pi}{4 \tau} \left(1+ 8 b_{\mathcal{N}=4}^2\right) }~.
\end{align}
Since with periodic boundary conditions $\mathcal{N}=4$ SYM theory has no Hagedorn instabilities along the line $\arg \tau = \pi/2$, the $\arg \tau \to \pi/2$ and $|\tau| \to 0$ limits commute. Setting $\arg \tau = \pi/2$ we thus obtain Eq.~\eqref{eq:2DbehaviorDetailed} with 
\begin{align}
\sigma_{\mathcal{N}=4} = - \frac{\pi}{4} \left(1 +8 b_{\mathcal{N}=4}^2\right).
\end{align}

%%%%%%%%%%%%%%
\begin{figure}[ht]
\centering
\includegraphics[width=\textwidth]{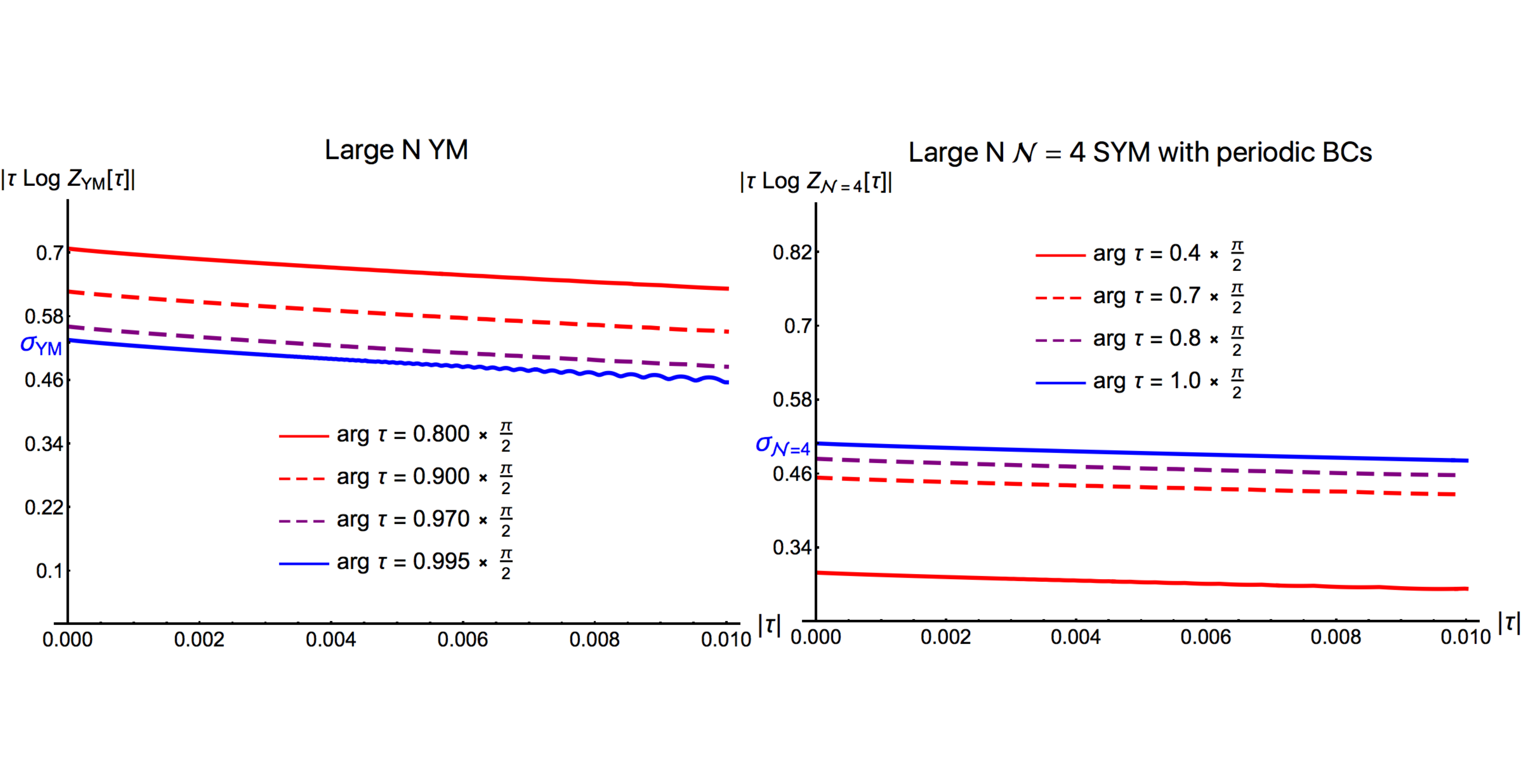}
\caption{The small-$|\tau|$ behavior of the confining-phase partition function of pure YM theory (left) and $\mathcal{N}=4$ SYM theory with periodic fermion boundary conditions (right), plotted as a function of $\arg \tau$ as $\arg \tau\to \pi/2$.} \label{fig:smallLYM}
\end{figure}
%%%%%%%%%%%%%%

Now let us consider pure YM theory. To calculate the small-$|\tau|$ behavior of $Z_{\rm YM}$ we need to know the behavior of $\eta(\tau), \Th{1/2}{b+1/2}(\tau),\Th{1/2}{0}(\tau)$ at small $|\tau|$. Here we defined $b = b(n_s = 0) = \frac{1}{2\pi}\cos^{-1}(2) \approx 0.2 i$. The small-$|\tau|$ behavior of $\eta(\tau)$ was already discussed above, while it is easy to see that
\begin{align}
\lim_{|\tau|\to 0} \Th{1/2}{0}(\tau) \sim (-i\tau)^{-1/2} .
\end{align}
The subtlety comes in the small-$|\tau|$ behavior of $\Th{1/2}{b+1/2}(\tau)$, which, using the $S$-transformation rule and the product representation of the $\Th{\alpha}{\beta}$ functions,  can be shown to take the form
\begin{align}
\lim_{|\tau|\to 0} 
\Th{1/2}{b+1/2}(\tau ) \sim (-i\tau)^{-1/2} e^{-\frac{i\pi}{2} (1-2b)} e^{-\frac{i\pi}{4 \tau} (3+2b)^2} 
\left[-1 + e^{-\frac{2\pi i b}{\tau}}\right]\left[-1 + e^{-\frac{2\pi i (b+1)}{\tau}}\right]~.
\end{align}
Putting the asymptotics together, we find that
\begin{align}
\lim_{|\tau|\to 0} 
Z_{\rm YM}(\tau)  \sim \frac{2 \sqrt{2}  e^{\frac{i \pi (8 b (b+3)+15)}{8 \tau }} \sin (\pi b)}{(-i \tau )^{3/2} \left(1-e^{\frac{2 i \pi b}{\tau }}\right) \left(-1+e^{\frac{2 i \pi (b+1)}{\tau }}\right)}
\label{eq:YMsmallL}
\end{align}
Since $b$ is pure imaginary, the factor in the denominator oscillates when $\arg \tau$ approaches $\pi/2$ and has zeroes when $\arg \tau = \pi/2$, so that Eq.~\eqref{eq:YMsmallL} has a sequence of poles along $\arg \tau = \pi/2$, with an accumulation point at $\tau = 0$. These are simply the modular $S$-transformation images of the Hagedorn singularities of $Z_{\rm YM}(\tau)$. As a result, the small-$|\tau|$ and $\arg \tau\to \pi/2$ limits do not commute, because it does not make sense to ask to approach $|\tau|=0$ along $\arg \tau = \pi/2$ using the confined-phase partition function. As already explained above, we take the small-$|\tau|$ limit before taking the $\arg \tau \to \pi/2$ limit, so that the confined-phase partition function remains well defined. In this limit, the Yang-Mills partition function behaves as
\begin{align}
\lim_{\arg \tau \to \pi/2} \left[ \lim_{|\tau| \to 0} \log Z_{\rm YM}(\tau)\right] \sim -\frac{\pi}{8} \left(1- 8 b^2\right) \frac{2\pi R}{\beta } \label{fixb}
\end{align}
so that
\begin{align}
\sigma_{\rm YM} = \frac{\pi}{8} \left(1- 8 b^2\right).
\end{align}

The calculations in these two examples can be performed for arbitrary $n_f, n_s$, and we find that Eq.~\eqref{eq:2DbehaviorDetailed} holds for all confined-phase large-$N$ adjoint-matter gauge theories on $S^3 \times S^1$ in the $\lambda \to 0$ limit. As we emphasized at the beginning of this subsection, this means that the coefficient $\sigma$ of the $\beta^{-3}$ term in the small-$\beta$ expansion of the confined-phase partition function vanishes. This cancellation is enforced by the modular symmetries of the confined-phase spectrum of large-$N$ theories.

%%%%%%%%%%%%%%%%%%
\subsection{2D CFT interpretation}
\label{sec:CFTinterpretation}
%%%%%%%%%%%%%%%%%%
The preceding two sections illustrated that the modular properties of the 4D confined-phase partition functions cause these partition functions to behave as if they correspond to two-dimensional CFTs. In this section, we shall make this 4D-2D connection sharper. Specifically, we shall argue that $Z_{\rm 4D}(\tau) = Z_{\rm 2D}(\tau)$, where $Z_{\rm 4D}$ are the confined-phase gauge theory partition functions and $Z_{\rm 2D}(\tau)$ are chiral partition functions of 2D CFTs. This shows that the 4D and 2D theories have coinciding spectra.

To show this connection we will simply exhibit 2D CFTs whose chiral partition functions coincide with gauge-theory partition functions. Of course, two quantum field theories can have coincident partition functions while having distinct correlation functions. Given just the spectral data, it is thus impossible to uniquely determine a 2D CFT associated with a specific 4D theory. The specific 4D-2D relations we propose below are therefore to be considered `proofs of principle' that large-$N$ gauge theories are indeed isospectral to 2D CFTs.

It would be very interesting to understand whether there is a large-$N$ 4D-2D equivalence for correlation functions and not merely for spectra. If such a mapping between generating functionals of 2D and 4D theories were to exist, it would presumably uniquely determine the 2D theories  appearing in the 4D-2D relation.  An exploration of this fascinating and challenging question is outside the scope of the present paper.

%%%%%%%%%%%%%
\subsubsection{Theories with $n_f = 0$ and arbitrary $n_s$}
\label{sec:CFTInterpretationBosons}
%%%%%%%%%%%%%

We begin by considering large-$N$ theories with $n_s$ adjoint scalars and no fermions. As we recall, these theories have partition functions given by 
\begin{align}
\label{eq:repeatBosonicPF}
Z^{\rm bosonic}_{\rm 4D}(\tau; n_s) &= \eta(\tau)^2 \cdot \eta(\tau)^2 \cdot \frac{1}{ \eta(\tau)} \cdot \frac{2 i \sin(\pi b_{n_s}) e^{-i\pi b_{n_s}} \eta(\tau)}{\Th{1/2}{b_{n_s}+1/2}(\tau)} \cdot \left[\frac{2\eta(\tau)}{\Th{1/2}{0}(\tau)}\right]^{1/2} \, \\
&= 1 + n_s q + (2+n_s)(3+n_s) q^2 + ... \,. \nonumber
\end{align}
We now show that these partition functions coincide with the chiral partition functions of a particular 2D CFT, as advertised in Eq.~\eqref{eq:TheClaim}.

To see this, we first recall that the $c=-26$ $bc$-ghost CFT has the chiral partition function
\begin{align}
 \eta(\tau)^2 = q^{1/12}\left(1 - 2 q- q^2 + ... \right) .
\end{align}
By `chiral' we mean that this partition function tallies contributions from, e.g., right-moving modes and lacks contributions from left-moving modes. Next, the chiral partition function of a $c=1$ non-compact free scalar CFT is given by
\begin{align}
\frac{1}{\eta(\tau)} = q^{-1/24}\left(1 + q + 2q^2 + 3 q^3 + 5 q^4 + ... \right).
\end{align}
We also observe that a $c=1$ scalar field with R-NS boundary conditions (that is, a scalar field which acquires a phase of $-1$ going around the thermal circle but which is  periodic along the spatial direction) has the chiral partition function~\cite{DiFrancesco:1997nk}
\begin{align}
\left[\frac{2\eta(\tau)}{\Th{1/2}{0}(\tau)}\right]^{1/2} = q^{-1/24} \left(1 - q - q^3 + ... \right).
\end{align}
Together, these observations account for four of the five factors in Eq.~\eqref{eq:repeatBosonicPF}. However, writing $z = e^{2\pi i b_{n_s}}$, we see that the remaining factor in Eq.~\eqref{eq:repeatBosonicPF} can be identified with the vacuum-sector chiral partition function 
\begin{align}
\frac{2 i \sin(\pi b_{n_s}) e^{-i\pi b_{n_s}} \eta(\tau)}{\Th{1/2}{b_{n_s}+1/2}(\tau)} &= q^{-1/12}\left[1+(z+z^{-1}) q + (1+z^{2}+z^{-2} +z +z^{-1}) q^2 + ... \right] 
\end{align}
of the $c=2$ bosonic $\beta \gamma$ ghost CFT~\cite{Ridout:2014oca}. This irrational logarithmic CFT has a $U(1)$ conserved charge and associated fugacity $z$.

Taking a direct product of these five CFTs, we then obtain a 2D CFT with a chiral partition function
\begin{align}
Z_{\textrm{2D CFT}}(\tau; n_s) =  \eta(\tau)^3 \, \frac{2 i \sin(\pi b_{n_s}) e^{-i\pi b_{n_s}} \eta(\tau)}{\Th{1/2}{b_{n_s}+1/2}(\tau)} \left[\frac{2\eta(\tau)}{\Th{1/2}{0}(\tau)}\right]^{1/2} \,. 
\end{align}
We thus have a special case of Eq.~\eqref{eq:TheClaim}, 
\begin{align}
Z^{\textrm{bosonic}}_{\textrm{4D}}(\tau; n_s) =  Z_{\textrm{2D CFT}}(\tau; n_s) ,
\end{align}
thereby establishing a spectral equivalence between a confined-phase large-$N$ 4D gauge theory and a 2D theory. 

Note that the parameter $n_s$ in the 4D gauge theory maps to a choice of fugacity for a conserved charge in the 2D theory. The resulting discrete values of the fugacities within the 2D theory have some remarkable properties. A generic chiral partition function $Z_{\rm 2D}$ of a 2D CFT can be schematically written as
\begin{align}
Z_{\rm 2D} = q^{\Delta} \sum_{n} \left(\sum_m c_{m,n} z^m\right)q^n,
\end{align} 
and one expects that $c_{m,n}$ must be integers. For generic values of $z$, there is no reason to expect that $\sum_m c_{m,n} z^m$ would be an integer. Yet for the particular values of $z$ relevant for the equivalence, $\sum_m c_{m,n} z^m$ is an integer. Moreover, the resulting coefficients of $q$ are non-negative. While the statement that the thermal partition function of a 4D bosonic gauge theory on $S^3 \times S^1$ has non-negative integer coefficients in its $q$-expansion is obvious from the perspective of the gauge theory, on the 2D CFT side working with $z = e^{2\pi i b_{n_s}}$ with $b_{n_s} = \frac{1}{2\pi} \cos^{-1}(2+n_s/2)$ corresponds to considering a set of extremely special points in the space of fugacities. 

It is tempting to speculate that these special points in the parameter space of the 2D CFT are associated with the emergence of enhanced symmetries. Indeed, large-$N$ gauge theories in the $\lambda \to 0$ limit are known to have an infinite tower of conserved higher-spin currents~\cite{Witten,Sundborg:2000wp}. Thus, it is possible that at these special points the Virasoro symmetry of the 2D CFT becomes enlarged to a $\mathcal{W}$-symmetry~\cite{Bouwknegt:1992wg}. This is an interesting point to explore in future work.

%%%%%%%%%%%%%
\subsubsection{Theories with fermionic matter fields}
\label{sec:CFTInterpretationFermions}
%%%%%%%%%%%%%
The 2D CFT interpretation for large-$N$ confining theories with generic adjoint matter proceeds in much the same way as for pure YM theory above. For concreteness, we start with the twisted partition function with generic $n_f, n_s$, which can be written as
\begin{align}
\tilde{Z}(\tau) = \prod_{\alpha = 1}^{3} \left[ 2 e^{-i \pi b_{\alpha}}\cos(\pi b_{\alpha}) \, \eta(\tau)^2 \frac{1}{\eta(\tau)}
\frac{ \eta(\tau) }{ \Th{1/2}{b_{\alpha}}(\tau)}
\frac{\eta(\tau)}{\Th{0}{b_{\alpha}}(\tau) } \right] 
\label{eq:twistedZmodularRepeat} \, .
\end{align}
Each of the factors in the finite-product expression above can be associated with the chiral partition function of a known 2D CFT, in a sector with given boundary conditions.

The factor of $\eta(\tau)^2 = q^{1/12}\left(1 - 2 q- q^2 + ... \right)$ coincides with the vacuum character of the $c = -26$ fermionic $bc$ ghost CFT.~ The factor of $1/\eta(\tau) = q^{-1/24}\left(1 + q + 2 q^2 + 3 q^3 + 5 q^4 + ... \right)$ coincides with the vacuum character of the non-compact $c=1$ free scalar CFT.~ Then one can observe that
\begin{align}
2 \frac{\cos(\pi b_{\alpha})}{e^{+i \pi b_{\alpha}}} \, \frac{ \eta(\tau) }{ \Th{1/2}{b_{\alpha}}(\tau)} 
&= 2i \frac{\sin(\pi [b_{\alpha}-\half])}{e^{\pi i (b_{\alpha} -\half)} } \frac{ \eta(\tau) }{ \Th{1/2}{(b_{\alpha}-1/2) +1/2}(\tau)} \nonumber \\
&= q^{-1/12}\big[1+(y_{\alpha}+y_{\alpha}^{-1}) q + (1+y_{\alpha}^{2}+y_{\alpha}^{-2} +y_{\alpha} +y_{\alpha}^{-1}) q^2 + ... \big] 
\end{align}
coincides the vacuum character of the $c = 2$ bosonic $\beta \gamma$ ghost CFT. Note that the flavor data of the gauge theory enters through the definition of the fugacity $y_{\alpha}$ of the $\beta \gamma$ CFT as $y_{\alpha} = e^{2\pi i (b_{\alpha}-1/2)}$, which relates to the fugacities defined elsewhere in this paper by $y_{\alpha} = -z_{\alpha}$. All of the factors mentioned thus far are invariant under $T:\tau \to \tau+1$ and have $q$-expansions involving only integer powers of $q = e^{2\pi i \tau}$. 

The remaining factor in Eq.~\eqref{eq:twistedZmodularRepeat} is different:
\begin{align}
\frac{\eta(\tau)}{\Th{0}{b_{\alpha}}(\tau) } = q^{1/24}\left[ 1 - (z_{\alpha}^{-1} + z_{\alpha})q^{1/2} + (1+ z_{\alpha}^{-2}+z_{\alpha}^{2}) q + ... \right]
\label{eq:cMinusOneGhost}
\end{align}
where $z_{\alpha} = e^{2\pi i b_{\alpha}}$.  However, this expression coincides with the chiral NS-R partition function of the $c = -1$ bosonic $\beta \gamma$ ghost CFT on the torus~\cite{Lesage:2002ch}. This shows that our general relation $Z_{\rm 4D} = Z_{\rm 2D}$ is actually satisfied for generic $n_s$ and $n_f$ in the case of large-$N$ confined-phase $S^3 \times S^1$ partition functions with periodic boundary conditions. This demonstration moreover supplies a concrete candidate for the 2D CFT entering Eq.~\eqref{eq:2DBehavior}. 

Eq.~\eqref{eq:cMinusOneGhost} is not a character function of the $c = -1$ ghost CFT because it is not a $T$-eigenstate. However, under the $T$ modular transformation we have $\eta(\tau)/\Th{0}{b_{\alpha}}(\tau) \to \eta(\tau)/\Th{0}{b_{\alpha}+1/2}(\tau)$. From the 4D perspective, this $T$-translation merely changes boundary conditions for the adjoint fermions from periodic to anti-periodic.   Thus we learn that the 4D thermal partition function can be interpreted with the same 2D product CFT as the twisted partition function, with the only change being a passage from the NS-R sector to the NS-NS sector in computing the contribution from the $c=-1$ bosonic ghost CFT.~  In Sect.~\ref{sec:GenericModularCompletion}, we will show that the modular orbits of 4D gauge theory include R-NS-type terms as well, in analogy to the 2D Ising model.

Finally, we can consider the $S^3 \times S^1$ confined-phase large-$N$ partition functions of supersymmetric theories, taking $\mathcal{N}=4$ SYM with periodic boundary conditions as a paradigmatic example:
\begin{align}
\tilde{Z}_{\kappa = 3}(\tau) &=
\frac{1}{\eta(\tau)}\left(\frac{\eta(\tau)}{ \Th{0}{\frac{1}{2}}(\tau)}\right)^2
\frac{ 2\cos(\pi b_{\kappa=3}) e^{- i \pi b_{\kappa=3} }\,\, \eta(\tau)^2}{ \Th{1/2}{b_{\kappa=3}}(\tau) \,\, \Th{0}{b_{\kappa=3}}(\tau)} \label{eq:paradigmatic} \, .
\end{align}
All of the ingredients appearing in the expression above have already been given a 2D CFT interpretation in our previous examples, except for $[\eta(\tau)/\Th{0}{1/2}(\tau)]^2 = q^{1/12}(1+ 4 q^{1/2} +10 q+...)$.  However this expression coincides with the chiral partition function of a $c=4$ CFT composed of two complex scalar fields with NS-R boundary conditions on the torus.  Similar remarks apply to the expression for supersymmetric theories with fewer adjoint matter supermultiplets, as well as to thermal partition functions. 

Thus, for all of the theories studied in this paper, we conclude that the large-$N$ gauge-theory partition functions coincide with chiral partition functions of 2D CFTs, as advertised in Eq.~\eqref{eq:TheClaim}. This then generalizes our previous results for pure Yang-Mills theory, as derived in Ref.~\cite{Basar:2015xda}.

%%%%%%%%%%%%%%%%%%%%%%%
\section{Characters and modular invariants}
\label{sec:ModularCompletion}
%%%%%%%%%%%%%%%%%%%%%%%
In Sect.~\ref{sec:ModularForms} we showed that the confined-phase partition functions of adjoint-matter large-$N$ gauge theories on $S^3 \times S^1$ can be written as combinations of modular forms.  Then, in Sect.~\ref{sec:CFTinterpretation}, we provided a 2D CFT interpretations of these 4D partition functions, thereby establishing our central claim in Eq.~\eqref{eq:TheClaim}. Our goal for this section is to gather information about the spectra of effective primary operator dimensions $h^{\rm (eff)}_i$ of the 2D CFTs that appear in Eq.~\eqref{eq:TheClaim}.  To this end, we will compute the diagonal modular invariants associated to the 2D CFTs appearing in Eq.~\eqref{eq:TheClaim}. This will allow us to compute the values of $h^{\rm (eff)}_i~(\textrm{mod} ~1)$. In all cases (aside from the semi-trivial case of the superconformal index), we shall find that $h^{\rm (eff)}_i$ form an infinite set with irrational values. These results are consistent with our matching of the chiral partition functions to 2D irrational CFTs in Sect.~\ref{sec:CFTinterpretation}.

%%%%%%%%%%%%%%%%%%%%
\subsection{Characters and modular invariants for theories with bosonic matter}
\label{sec:BosonicModularCompletion}
%%%%%%%%%%%%%%%%%%%%

In our earlier discussion of large-$N$ gauge theories with $n_s$ conformally-coupled massless adjoint scalar fields and no fermions, we found that the confined-phase partition functions take the form given in Eq.~\eqref{eq:BosonicZ} and hence have a clear modular structure. However, they are not modular invariant, and their 2D interpretation is in terms of a chiral sector of a 2D CFT. Except in the very special context of chiral 2D CFTs, modular invariance in a 2D CFT requires that  we include the contributions of both left and right-moving sectors and sum over these sectors in a way consistent with the modular symmetries. There can be more than one consistent way to stitch together the left and right-moving sectors, corresponding to the possibility of introducing orbifold projections. Here we shall consider the simplest modular invariant associated to Eq.~\eqref{eq:BosonicZ}, namely the diagonal modular invariant.

Given a `seed' chiral partition function $Z_{\rm seed}$ of modular weight $k$, the corresponding diagonal modular invariant $\mathcal{Z}_{\rm diagonal}$ can be formally defined as a sum over the modular images of $Z_{\rm seed}$:
\begin{align}
\mathcal{Z}_{\rm diagonal}(\tau) \overset{!}{=} (\im \tau)^{k}\sum_{\gamma \in SL(2, \mathbb{Z})} |Z_{\rm seed}(\gamma \cdot \tau)|^2.
\label{eq:FormalInvariant}
\end{align} 
We shall employ the symbol $\overset{!}{=}$ to emphasize that the right sides of such equations may require a regularization consistent with the modular symmetries in order to make the relation precise.   This will be discussed further below.   Note the factor of $(\im \tau)^{k}$ in Eq.~(\ref{eq:FormalInvariant}) can be thought of as the contribution to $\mathcal{Z}_{\rm diagonal}$ of the zero-mode excitations of the CFT (which are neither left- nor right-moving), and must be present for $\mathcal{Z}_{\rm diagonal}(\tau)$ to be fully modular invariant. Once we know the form of $\mathcal{Z}_{\rm diagonal}$ for a CFT, it will be straightforward to extract information about the corresponding primary operator spectrum.

In the most familiar cases, such as those involving the CFTs corresponding to the so-called ``minimal models'', there are an infinite number of elements $\gamma$ which map the seed term $Z_{\rm seed}$ to itself. This will happen if $Z_{\rm seed}$ is built from, e.g., Dedekind $\eta$ functions and Jacobi $\Th{a}{b}$ functions with rational characteristics $a, b \in \mathbb{Q}$. In such cases, the set of modular transformations $\gamma$ has a natural decomposition into equivalence classes, defined such that any two elements $\gamma_1, \gamma_2$ of $SL(2,\mathbb{Z})$ belong to the same equivalence class if they have the same action on $Z_{\rm seed}$, with  $Z_{\rm seed}(\gamma_1 \cdot \tau) = Z_{\rm seed}(\gamma_2 \cdot \tau)$.    This redundancy leads to a divergence in the naive expression in Eq.~\eqref{eq:FormalInvariant},  since the size of each equivalence class is generally infinite.  In such cases we must instead choose a single representative from each distinct equivalence class in defining $\mathcal{Z}_{\rm diagonal}$ in order to obtain a convergent version of Eq.~\eqref{eq:FormalInvariant}, and this may be considered to be a kind of regulator. However, in our case, $Z_{\rm seed} (\tau)= Z(\tau; n_s)$ contains a $\vartheta$ function with an \emph{irrational} characteristic.  This in turn implies that each element of $SL(2,\mathbb{Z})$ will have a unique action on $Z(\tau; n_s)$.  Consequently the sum in Eq.~\eqref{eq:FormalInvariant} will contain an infinite number of distinct terms, and we will not have to worry about splitting the modular orbit of $Z_{\rm seed}$ into equivalence classes and picking representatives.  Indeed, all of the terms in the modular orbit of $Z_{\rm seed}$ will be needed in order to construct the diagonal invariant.

To give an explicit description of the diagonal invariant, we construct a set of objects $\{Z_{m,n}\}$, where the indices $m, n$ are relatively prime integers, that have the following properties:
\begin{enumerate}
\item[(a)] Each element $Z_{m,n}$ is built out of a finite product of modular functions with modular weight $k = 3/2$.
\item[(b)] The element $Z_{m,n}$ reduces to the QFT partition function when $m = 0, n=1$ so that 
\begin{align}
Z_{0,1}(\tau; n_s) = Z_{\rm bosonic}(\tau; n_s)
\end{align}
\item[(c)] The set $\{Z_{m,n}\}$ is closed under the action of the modular group, in the sense that
\begin{align}
Z_{m, n}(-1/\tau; n_s) &= (-i \tau)^{3/2} s_{m,n} Z_{-n, m}(\tau; n_s) \nonumber\\
Z_{m, n}(\tau+1; n_s) &= t_{m,n} Z_{m, m+n}(\tau; n_s),
 \label{eq:TmnTActionGeneral}
\end{align}
where $s_{m,n}$ and $t_{m,n}$ are pure phases which do not depend on $\tau$, and there is a one-to-one mapping of the action of $SL(2,Z)$ on the argument $\tau$ to an action on the indices $m,n$ of the elements of the set $\{Z_{m,n}(\tau)\}$. This means that if we view the indices $m, n$ as the components of a column vector, then up to a factor of $(-i \tau)^{3/2}$ each element of $SL(2, Z)$ acts by matrix multiplication on this column vector.
\end{enumerate}
In the rest of this section, we will mostly use notation where the dependence of $Z_{m,n}$ on $n_s$ is suppressed, so that $Z_{m, n}(\tau; n_s)$ is abbreviated as $Z_{m, n}(\tau)$.

We define the elements of the set $\{Z_{m,n}(\tau)\}$ as
\begin{align}
Z_{m,n}(\tau) ~\equiv~ 
 \frac{- [2+n_s]^{1/2} \eta(\tau)^4}{ e^{i\pi n\, b_{n_s}}\Th{m\, b_{n_s} + 1/2}{n\, b_{n_s} + 1/2}(\tau)} \left[\frac{2\eta(\tau)}{\Th{P(m)/2}{P(n)/2}(\tau)}\right]^{1/2}, \label{Tb01}
\end{align}
where
\begin{align}
P(m) ~\equiv~ \half \left[1+(-1)^m\right] = \begin{cases} 1   &   m ~ {\rm even} \\  0   &   m ~ {\rm odd.} \\ \end{cases} \label{pF} 
\end{align}
With this definition of $Z_{m,n}(\tau)$, condition (a) is clearly satisfied, and so is condition (b), because by construction $Z_{0,1}(\tau) = Z_{\rm bosonic}(\tau)$.

We now observe that condition (c) is also satisfied. First, using the identities collected in Appendix~\ref{sec:AppendixConventions}, as well as the identity
\begin{align}
P(n) + P(m) + 1 =  P(m+n) ~~ ( {\rm mod} \, 2) \, , 
\label{pF2}
\end{align}
we find that the $S$-transformation rule is given by
\begin{align}
Z_{m,n}(-1/\tau; n_s) = (-i \tau)^{3/2} e^{2\pi i(m\, n\, b_{n_s}^2+1/4)} Z_{-n, m}(\tau; n_s)~. 
\end{align}
Since $b_{n_s}^2$ is a real number, this means that $s_{m,n} = e^{2\pi i(m\,n\,b_{n_s}^2+1/4)}$ is indeed a pure phase, verifying Eq.~\eqref{eq:TmnTActionGeneral}. Similar manipulations show that the $T$-transformation rule for $Z_{m,n}(\tau)$ is
\begin{align}
Z_{m,n}(\tau+1) =e^{\pi i \left\lbrace [1-P(m)]/8 - m^2 (b_{n_s})^2 \right\rbrace} Z_{m, n+m}~, 
\end{align}
so that $t_{m,n} = e^{\pi i \left\lbrace [1-P(m)]/8 - m^2 (b_{n_s})^2 \right\rbrace}$ is also a pure phase. (An explicit proof of this fact is contained within Appendix~\ref{sec:BosonicSimplifcations}.)

We now observe that the integers $(0,1)$ labeling the seed term are relatively prime. Likewise,  if $(m,n)$ are relatively prime, then so are $(-n, m)$ and $(m, m+n)$. This means that the modular orbit of the seed term is contained within the set of pairs of co-prime integers. In fact, any co-prime pair $(m,n)$  can be mapped back to $(0,1)$  by some element $M$ of $PSL(2,\mathbb{Z})$, so that the modular orbit requires \emph{all}\/ relatively prime pairs $(m,n)$. To see this, let $(m,n)$ be an arbitrary co-prime pair.  Our goal is then to solve for the matrix $M$ for which
\begin{align}
M \cdot \begin{pmatrix} {m} \\ {n} \end{pmatrix} 
\equiv \begin{pmatrix} a & b\\ c &d \end{pmatrix} \cdot \begin{pmatrix} {m}\\ {n} \end{pmatrix}
= \begin{pmatrix} a m + b n\\ c m + d n \end{pmatrix} 
= \begin{pmatrix} 0\\ 1 \end{pmatrix}.
\end{align}
To solve for $M$, we first observe by B\'ezout's lemma that for any coprime $(m,n)$ there exist integers $(c,d)$ such that $c m + d n = 1$. However, once suitable integers $(c,d)$ are chosen, we must also ensure that $a m + b n = 0$. This is solved by setting $a = n k, b = -m k$, for $k \in \mathbb{Z}$. Hence $a,b,c,d$ are determined up to an integer $k$. The condition $\det M = 1$ then implies $1= [n k d -(-m k) c]$, and this fixes $k=1$.  Thus $M$ is indeed an element of $PSL(2,Z)$. This then completes the demonstration of statement (c): the set $\{ Z_{m,n}(\tau)| m\perp n \in \mathbb{Z}\}$ is closed under modular-group transformations, where the notation $m \perp n$ indicates that $m$ and $n$ are coprime. 

If $b_{n_s}$ had been rational, our  verification of condition (c) above would have gone through without change. The only difference would have involved the structure of the modular orbits. If $b_{n_s}$ had been rational, after some finite number of applications of $S$ and $T$ to $Z_{0,1}$ we would have returned to $Z_{0,1}$. We would then have needed to break the set $\{Z_{m,n}, m\perp n\}$ into equivalence classes and take a single representative from each equivalence class.  This would have resulted in a {\it finite}\/ modular orbit. However, as we already mentioned above, $b_{n_s}$ is irrational for all non-negative integers $n_s$. This means that the modular orbit of $Z_{0,1}$ is {\it infinite}\/-dimensional, and each distinct pair of coprime integers $(m,n)$ is associated with a distinct element $Z_{m,n}$ of the orbit.

\begin{figure*}[h]
\centering
\includegraphics[width=0.45\textwidth]{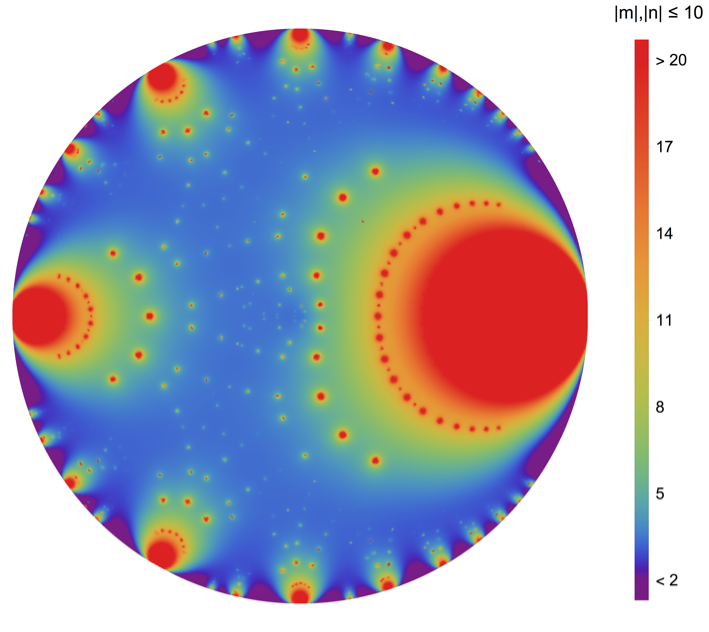}
\includegraphics[width=0.45\textwidth]{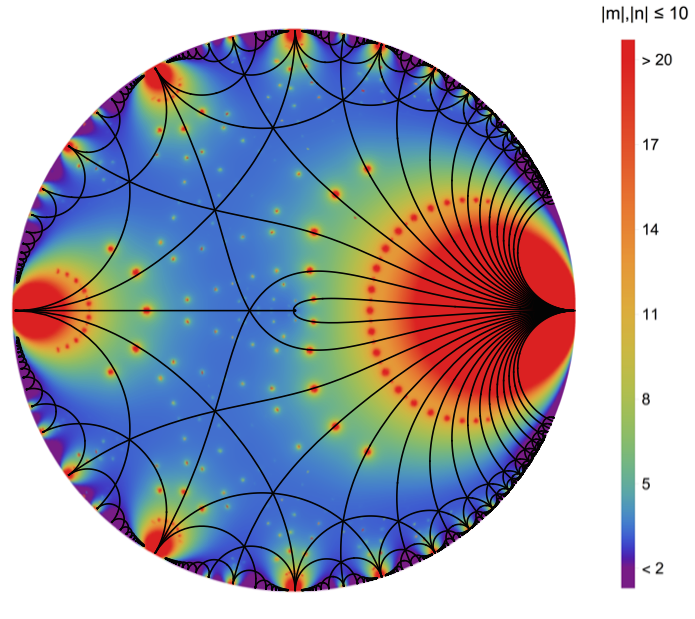}
\caption{{\it Left:}\/ Values of $\mathcal{Z}_{\rm diagonal}(\tau; n_s = 0 )$, corresponding to pure Yang-Mills theory, evaluated with a cutoff $|m|,|n|< 10$ on the sum in Eq.~\eqref{eq:bModular1} and plotted within the unit $q$-disk. {\it Right:}\/ The overlay of the same plot with the tessellations generated by the modular transformations.}
\label{fig:Tconvergence}
\end{figure*}

Armed with these observations, we can write down the minimal modular completion of the seed term:
\begin{align}
\mathcal{Z}_{\rm diagonal}(\tau; n_s) = (\im \tau)^{3/2} \sum_{m \perp n} |Z_{m,n}(\tau; n_s) |^2 \, . 
\label{eq:bModular1}
\end{align}
One may wonder whether the infinite sum over $m$ and $n$ converges for $\tau \in \mathbb{H}$ and non-negative integers $n_s$. Our numerical evidence suggests that the sum converges at generic points in the complex-$\tau$ half-plane $\mathbb{H}$, except for an isolated set of points associated to Hagedorn singularities. The numerical values of $\mathcal{Z}_{\rm diagonal}(\tau; n_s = 0)$ as a function of a cutoff on the sum over $Z_{m,n}$ are illustrated in Fig~\ref{fig:Tconvergence}. 

The fact that $\mathcal{Z}_{\rm diagonal}(\tau; n_s)$ includes the seed term $Z_{0,1}$, is modular invariant, and composed of absolute values of $Z_{m,n}$ implies that $\mathcal{Z}_{\rm diagonal}(\tau; n_s)$ has many more Hagedorn singularities than $Z_{0,1}$. Indeed, we already know that $Z_{0,1}(\tau; n_s)$ has an isolated set of Hagedorn singularities, for instance along the interval $(0,1)$ of the $q$-disk, with an accumulation point at $q = 1$. But the $S$-modular image of $Z_{1,0}$, which is of course included in $\mathcal{Z}_{\rm diagonal}(\tau; n_s)$, must then have a set of Hagedorn singularities on $q \in (0,1)$ with an accumulation point at $q = 0$. In this way, the $S$- and $T$- transformations produce $SL(2,\mathbb{Z})$ images of the Hagedorn singularities of the seed term $Z_{0,1}$ throughout the upper half-plane. 

The singularities of $Z_{m,n}(q)$ are identified with the zeroes of the theta function in the denominator of Eq.~\eqref{Tb01}. Using the  triple-product representations in Eq.~\eqref{eq:triple_product}, we see that the singularities are simple poles located at
\begin{equation}
q^{(m,n)}_{\star}= \exp \left(2\pi i {n i |b| +k\over -m i |b| +l} \right)\,,\quad k,l\in {\mathbb Z}
\end{equation}  
with the restriction that $q^{(m,n)}_{\star}$ lies inside the unit circle. In writing this expression we used the fact that $b = \frac{1}{2\pi} \cos^{-1}(2+n_s/2) = + i |b|$. In the complex $\tau$-plane these singularities are mapped to 
\begin{align}
\tau^{(m,n)}_{\star}= {n i |b| +k\over -m i |b| +l} \, ,~~\quad k,l\in {\mathbb Z}\,.
\label{eq:tau_singularities}
\end{align}
This expression is expected in the following sense. The seed partition function $Z_{0,1}$ has Hagedorn singularities at $\tau=i |b|/q$ with $q\in{\mathbb Z}^+$. General modular transformations map this set of seed singularities to the set in Eq.~\eqref{eq:tau_singularities}. It is relatively simple to show that since $(m,n)$ are relatively prime, no $Z_{m,n}(\tau)$ share poles.

Moving forward, we would like to extract the spectrum of (effective) conformal dimensions in the full modular 2D CFT. General 2D CFT considerations indicate that the eigenvalues of the modular $T: ~ \tau \to \tau+1$ operator encode data concerning the spectrum of primary operators of the CFT.~ If the scaling dimensions are real, the eigenvalues of $T$ will be pure phases, and the set of these eigenvalues can be written as
\begin{align}
\{e^{2\pi i\, h^{\rm (eff)}_{k}}\}
\end{align}
where $k$ is an index parametrizing the elements of the set. In the simplest examples, such as minimal-model CFTs, $k$ takes a finite number of values.  However,  in the more generic case of irrational CFTs, $k$ may be drawn from an uncountably infinite set. The spectrum of primary operators is encoded in the values of $h^{\rm (eff)}_{k}$, which are related to the scaling dimensions of the primary operators $h_k$ via 
\begin{align}
h^{\rm (eff)}_{k} = h_{k} - c/24 ~~\mathrm{(mod~1)}~.
\end{align}
Here $c$ is the central charge of the 2D CFT.~ So if we are able to compute the eigenvalues of $T$, we essentially determine $h^{\rm (eff)}_{k} ~\textrm{(mod~1)} $:
\begin{align}
\textrm{Eigenvalues of $\,T$} ~~ \Longleftrightarrow ~~ h^{\rm (eff)}_{k} ~\textrm{(mod~1)}~ . 
\end{align}
Note that without further assumptions, the eigenvalues of $T$ allow us to determine $ h^{\rm (eff)}_{k} $ only up to shifts by integers;  moreover the calculation cannot determine $h_k$ and $c$ separately. If one were to further assume that the underlying 2D CFT is unitary, then one would know that $\mathrm{min}\{h_{k}\} = 0$. The lowest value of $h^{\rm (eff)}_{k}$ would then yield $c$ on its own, which would in turn allow us to determine the spectrum of values of $\{ h_k\}$, up to integer shifts. Unfortunately, there is ultimately no compelling reason to expect our CFTs to be unitary. Indeed, such an assumption would not be consistent with our proposed identification of these 2D CFTs 
as containing logarithmic sectors, as outlined in Sect.~\ref{sec:CFTinterpretation}.

Our task is now to construct eigenstates of $T: \tau \to \tau+1$. Since $T$ is a discrete translation operator in the complex $\tau$-plane, the construction of $T$-eigenstates closely parallels the construction of Bloch-wave eigenstates for particles in periodic potentials. To write explicit expressions for the eigenstates, we first observe that the $T$-transformation leaves the first index of $Z_{m,n}$ invariant, $T: m \to m$, while it acts on the second index as $T: n \to n+m$. But any $n$ which is coprime to $m$ can be written as $k m +\ell$ for some $k \in \mathbb{Z}$ and an integer $\ell$ satisfying $0 \le \ell < |m|$. Thus, given a fixed index $m$, the set $\lbrace Z_{m,n}\rbrace $ can be decomposed into $\phi(m)$ `blocks', parametrized by $\ell$, which do not mix with each other under the action of $T$. Given this observation, it is then easy to see that the eigenstates are built from linear combinations of $\lbrace Z_{m,n}\rbrace$ which are labeled by $m$, $\ell$, and a Bloch `angle' $\alpha$. Explicitly, we find that
\begin{align}
\chi_{m,\ell,\alpha}(\tau)=
\sum_{k =-\infty}^{+\infty} e^{2\pi i \alpha k}Z_{m, k m+ \ell}(\tau) \,,
\label{eq:bChar2}
\end{align}
are eigenstates of $T$, and obey the relation
\begin{align}
\chi_{m,\ell,\alpha}(\tau+1) = e^{2\pi i h^{\rm (eff)}_{m, \ell,\alpha} } \chi_{m, \ell, \alpha}(\tau)
\end{align}
where
\begin{align}
h^{\rm (eff)}_{m, \ell,\alpha} ~~ (\textrm{mod 1}) = \frac{1}{2} \left(\frac{1-P(m)}{8} + m^2|b_{n_s}|^2 \right) - \alpha .
\label{eq:heffBosonic}
\end{align}
The set $\{\chi_{m, \ell, \alpha}(\tau)\}$ is a complete basis for the eigenstates of $T$, as can be checked by verifying that summing $|\chi_{m, \ell, \alpha}(\tau)|^2$ over the labels $m, \ell,\alpha$ reproduces the diagonal modular invariant:
\begin{align}
Z_{\rm diagonal}(\tau; n_s) = ({\rm Im}\, \tau)^{3/2}\, 
\sum_{m \in \mathbb{Z}}\, 
\sum_{\substack{ 0\leq \ell < |m|\\ \ell\perp m}} \, 
\int_{0}^{1} d\alpha \, |\chi_{m,\ell, \alpha}|^2 . \label{eq:bModular2}
\end{align}

This confirms that the quantities in Eq.~\eqref{eq:heffBosonic} are the set of scaling dimensions of primary operators (mod~$1$) of any 2D CFT which is isospectral to 4D large-$N$ gauge theory with $n_s$ adjoint conformally-coupled massless scalar fields in the confined phase on $S^3 \times S^1$. Note that $h^{\rm (eff)}_{m, \ell,\alpha} ~ (\textrm{mod~1})$ does not depend on $\ell$, but does depend on $\alpha$, which is a continuous variable. This quantity also depends on $|b|^2$, which is irrational. Moreover, the value of $\alpha$ is independent of $b$. Thus the scaling dimensions $h^{\rm (eff)}$ are irrational, and consequently any 2D CFT which is isospectral to this class of confining large-$N$ 4D gauge theories must be an irrational CFT~\cite{Anderson:1987ge,Vafa:1988ag}. This result is consistent with our identification of the candidate 2D CFTs in Sect.~\ref{sec:CFTInterpretationBosons}.

%%%%%%%%%%%%%%%%%%%
%%%%%%%%%%%%%%%%%%%
\subsection{Characters and modular invariants for theories with fermionic matter}
\label{sec:GenericModularCompletion}
%%%%%%%%%%%%%%%%%%%
%%%%%%%%%%%%%%%%%%%

Gauge theories with fermionic matter have more complicated modular structures than those with purely bosonic matter, for the following reason. First, there are two different types of boundary conditions for fermions on the $S^1$, periodic and anti-periodic. Second, $T$-translations exchange these boundary conditions, because fermionic states have half-integral energies in units of $1/R$, and $T$ maps $q^{n/2}$ to $(-1)^{n} q^{n/2}$. In other words, the modular completions of 4D large-$N$ gauge theories with fermionic matter content necessarily includes both periodic and anti-periodic boundary conditions. 

For simplicity, we will focus our discussion on generic matter content $n_f$ and $n_s$. While supersymmetric matter content simplifies the individual modular structure of the seed terms, it does not significantly alter the general form of the orbits when the modular parameter is identified as $q = e^{-\beta/R} = e^{2\pi i \tau}$. (Things are different if one defines the modular parameter via $e^{-\beta/(2R)} = e^{2\pi i \tilde{\tau}}$, as discussed in Appendix~\ref{sec:OtherModular}.) 

As we will focus our attention on the non-supersymmetric cases, we directly study the modular orbits of the expressions in Eqs.~\eqref{twistZ0} and \eqref{thermZ0}. Our goal is to find the general class of modular objects that naturally include these expressions.  The construction is very similar to that in Sect.~\ref{sec:BosonicModularCompletion}, and just as in that section it is helpful to focus on the parts of these expressions which contain theta functions with the non-trivial, i.e., complex and transcendental, characteristics. Hence, we focus on the modular orbits of
\begin{align}
T^{A}_{0,1}(\tau)&=\prod_{\alpha = 1}^{3} \frac{ e^{-i \pi b_{\alpha}} \, \eta(\tau) }{ \Th{1/2}{b_{\alpha}}(\tau)} \frac{ \eta(\tau) }{\Th{0}{b_{\alpha}}(\tau) } \,  \nonumber \\
T^{B}_{0,1}(\tau)&=\prod_{\alpha = 1}^{3} \frac{e^{-i \pi b_{\alpha}} \, \eta(\tau) }{ \Th{1/2}{b_{\alpha}}(\tau)} \frac{ \eta(\tau) }{\Th{0}{b_{\alpha}+\half}(\tau) } \, , \label{eq:Zb1}
\end{align}
where $T^{A}_{0,1}(\tau)$ originates from the twisted partition function in Eq.~\eqref{twistZ0} while $T^B_{0,1}(\tau)$ originates from the thermal partition function in Eq.~\eqref{thermZ0}. To do this, we begin by defining \emph{three} infinite families of terms $T^A_{m,n}(\tau), T^B_{m,n}(\tau)$, and $T^C_{m,n}(\tau)$:
\begin{align}
T^{A}_{m,n}(\tau)&=\prod_{\alpha = 1}^{3} 
\frac{e^{-i \pi n P(m) b_{\alpha}} \, \eta(\tau) }{\Th{mb_{\alpha} + P(m)/2}{nb_{\alpha} + P(n)/2}(\tau)} 
\frac{e^{-i \pi n \Bar{P}(m) b_{\alpha}} \, \eta(\tau) }{ \Th{mb_{\alpha} + \bar{P}(m)/2}{nb_{\alpha} + \bar{P}(n)/2}(\tau) } \, \nonumber\\
T^{B}_{m,n}(\tau)&=\prod_{\alpha = 1}^{3} 
\frac{e^{-i \pi n P(m) b_{\alpha}} \, \eta(\tau) }{\Th{mb_{\alpha} + P(m)/2}{nb_{\alpha} + P(n)/2}(\tau)} 
\frac{e^{-i \pi n \Bar{P}(m) b_{\alpha}} \, \eta(\tau) }{ \Th{mb_{\alpha} + \bar{P}(m)/2}{nb_{\alpha} + P(n)/2}(\tau) } \,\nonumber\\
T^{C}_{m,n}(\tau)&=\prod_{\alpha = 1}^{3} 
\frac{e^{-i \pi n P(m) b_{\alpha}} \, \eta(\tau) }{\Th{mb_{\alpha} + P(m)/2}{nb_{\alpha} + P(n)/2}(\tau)} 
\frac{e^{-i \pi n P(m) b_{\alpha}} \, \eta(\tau) }{ \Th{mb_{\alpha} + P(m)/2}{nb_{\alpha} + \bar{P}(n)/2}(\tau) } \, .
\label{eq:Zc2}
\end{align}
where $m$ and $n$ run over the full set of relatively prime integers and where, in close analogy with the function$P(m)$ defined in Eq.~(\ref{pF}),  we have now additionally defined
\begin{align}
\bar{P}(m) ~\equiv~ \frac{1}{2} [1-(-1)^{m}] = \begin{cases} 0   &   m ~ {\rm even} \\ 1   &   m ~ {\rm odd.} \\ \end{cases} \label{pFbar}  
\end{align}

Under the $T$ modular transformation, we find
\begin{align}
{T}&: ~~~ T^A_{m,n}(\tau) \to T^B_{m,n+m}(\tau) \, 
\prod_{\alpha = 1}^{3} {\rm exp}\left[i\pi \left\{ \left(m^2 b_{\alpha}^2 + \frac{P(m)}{4} \right) 
+ \left(m^2 b_{\alpha}^2 + \frac{\bar{P}(m)}{4} \right) \right\} \right]  \nonumber \\
{T}&: ~~~T^B_{m,n}(\tau) \to T^A_{m,n+m}(\tau) \, 
\prod_{\alpha = 1}^{3} {\rm exp}\left[i\pi \left\{ \left(m^2 b_{\alpha}^2 + \frac{P(m)}{4} \right) 
+ \left(m^2 b_{\alpha}^2 + \frac{\bar{P}(m)}{4} \right) \right\} \right]  \nonumber \\
{T}&: ~~~T^C_{m,n}(\tau) \to T^C_{m,n+m}(\tau) \, 
\prod_{\alpha = 1}^{3} {\rm exp}\left[i\pi \left\{ \left(m^2 b_{\alpha}^2 + \frac{P(m)}{4} \right) 
+ \left(m^2 b_{\alpha}^2 + \frac{P(m)}{4} \right) \right\} \right]     \, . \label{TzC} 
\end{align}
The proof of the results in Eq.~\eqref{TzC} depends on the identities
\begin{align}
& \bar{P}(a) + P(b) + 1 = \bar{P}(a+b) \,\,\, ({\rm mod} \,\, 2) \nonumber \\
& \bar{P}(a) + \bar{P}(b) + 1 = P(a+b) \,\,\, ({\rm mod} \,\, 2) \nonumber \\
& P(a) + \bar{P}(b) + 1 = \bar{P}(a+b) \,\,\, ({\rm mod} \,\, 2) ~ 
\label{l4} 
\end{align}
in addition to the identity in Eq.~(\ref{pF2}). By contrast, the $S$ modular transformation shuffles the characters in a slightly different way:
\begin{align}
{S}&:~~~ T^A_{m,n}(\tau) \to T^A_{-n,m}(\tau) \, 
\prod_{\alpha = 1}^{3} {\rm exp}\left[-2 i \pi \left\{ \left(m n b_{\alpha}^2 \right) 
+ \left(m n b_{\alpha}^2 + \frac{\bar{P}(m)\bar{P}(n)}{4} \right) \right\} \right] \nonumber\\
{S}&:~~~ T^B_{m,n}(\tau) \to T^C_{-n,m}(\tau) \, 
\prod_{\alpha = 1}^{3} {\rm exp}\left[-2 i \pi \left\{ \left(m n b_{\alpha}^2 \right) 
+ \left(m n b_{\alpha}^2 + \frac{\bar{P}(m)P(n)}{4} \right) \right\} \right]     \nonumber\\
{S}&:~~~ T^C_{m,n}(\tau) \to T^B_{-n,m}(\tau) \, 
\prod_{\alpha = 1}^{3} {\rm exp}\left[-2 i \pi \left\{ \left(m n b_{\alpha}^2 \right) 
+ \left(m n b_{\alpha}^2 + \frac{P(m)\bar{P}(n)}{4} \right) \right\} \right]~.      
\label{SzC} 
\end{align}
We thus see that the objects $T^A_{m,n}(\tau)$, $T^{B}_{m,n}(\tau)$, and $T^C_{m,n}(\tau)$ in Eq.~\eqref{eq:Zc2} map into each other under the $S$- and $T$-transformations in Eqs.~\eqref{TzC} and~\eqref{SzC}. Indeed, their $S$- and $T$-transformation rules are exactly the same as those experienced by the 2D massless fermion characters with NS-R, R-NS, and NS-NS boundary conditions, respectively.

Importantly, the integers $(m,n)$ that characterize a given modular image of a seed term behave in the same way for bosonic and fermionic large-$N$ gauge theories. In other words, we again find that $T: (m,n) \to (m,n+m)$ and $S: (m,n) \to (n, -m)$ in Eqs.~\eqref{TzC} and~\eqref{SzC} for generic $(N_f,N_s)$, just as we found previously in Eq.~\eqref{eq:TmnTActionGeneral} for the purely bosonic case when $N_f = 0$ with arbitrary (positive, integer) $N_s$.  Moreover, we have numerically verified that the phases under $S$- and $T$-transformations in Eqs.~\eqref{TzC} and~\eqref{SzC} are indeed pure phases, with modulus one. Thus, we can recycle the logic from the purely bosonic case to conclude that modular images of the seed term exist for every pair of coprime integers $(m,n)$. The complete modular invariant for generic $(N_f,N_s)$ is thus the sum of the squared moduli of the modular images of the seed terms:
\begin{align}
 Z_{\rm diagonal}(\tau; n_f, n_s) = ({\rm Im}\, \tau)^{3/2}\, \sum_{k \in \{{\rm A,\, B,\, C}\}} \sum_{m \perp n} |Z^{k}_{m,n}(\tau)|^2 \, , \label{xGeneric}
\end{align}
where $Z^{k}_{m,n}(\tau; n_f,n_s)$ is simply $T^{k}_{m,n}(\tau; n_f, n_s) \prod_{\alpha = 1}^3 2 \cos(\pi b_{\alpha}) \eta(\tau)$, as is needed to match the seed terms in Eqs.~\eqref{twistZ0} and~\eqref{thermZ0}.

The decomposition of the expression in Eq.~\eqref{xGeneric} into $T$-eigenstates goes through in the same manner as for purely bosonic theories, with one structural difference. The difference arises because there are now three sets of terms in the modular orbit, and $T$-translations map elements of $\{T^A_{m,n}(\tau)\}$ into elements of $\{T^B_{m,n}(\tau)\}$ and vice versa, while they map elements of $\{T^C_{m,n}(\tau)\}$ amongst themselves. As a result, we find that the $T$-eigenstates can be written as: 
\begin{align}
&\chi^{\rm I}_{m,l,\alpha}(\tau) = \sum_{k \in \mathbb{Z}} e^{2 \pi i \alpha k}\frac{1}{\sqrt{2}} \, Z^C_{m,k\cdot m + \ell}(\tau)  \nonumber\\
&\chi^{\rm II}_{m,l,\alpha}(\tau) = \sum_{k \in \mathbb{Z}} e^{2 \pi i \alpha k}\frac{1}{\sqrt{2}}\big[Z^A_{m,k\cdot m + \ell}(\tau) + Z^B_{m,k\cdot m + \ell}(\tau)\big] \nonumber\\
&\chi^{\rm III}_{m,l,\alpha}(\tau) = \sum_{k \in \mathbb{Z}} e^{2 \pi i \alpha k}\frac{1}{\sqrt{2}}\big[Z^A_{m,k\cdot m + \ell}(\tau) - Z^B_{m,k\cdot m + \ell}(\tau)\big]~.
\end{align}
This complete, orthonormal basis of $T$-eigenstates allows us to rewrite the modular completion of the 4D QFT partition function as
\begin{align}
Z_{\rm diagonal}(\tau; n_f, n_s) = ({\rm Im}\, \tau)^{3/2}\, \sum_{k \in \{{\rm I,\, II,\, III}\}} \sum_{m \in \mathbb{Z}}\, \sum_{\substack{ 0\leq \ell < |m|\\ \ell\perp m}} \, \int_{0}^{1} d\alpha \, |\chi^{k}_{m,\ell, \alpha}|^2 \, . \label{xGeneric}
\end{align}
The structural parallels between this expression and the expressions we found in purely bosonic theories can be used to verify that the $T$-eigenvalue phases $h^{\rm (eff)}_k$ are drawn from a continuous set and are irrational. Consequently, any 2D CFT which is isospectral to large-$N$ confined-phase gauge theories with fermions must be irrational. This is again consistent with our identification of the candidate 2D CFTs in Sect.~\ref{sec:CFTInterpretationFermions}.

%%%%%%%%%%%%%%%%%%%%
\section{Discussion}  
\label{sec:Trees} 
%%%%%%%%%%%%%%%%%%%%

Our goal in this work has been to understand whether there may be interesting emergent symmetries organizing the spectra of large-$N$ confining theories. We explored this question in the context of large-$N$ gauge theories with massless matter on $S^3_R \times S^1_{\beta}$, and used $R\Lambda$ as a control parameter in order to restrict our attention to the regime $R\Lambda \to 0$, where these theories become solvable at large $N$. 

We found that in this setting the confined-phase partition functions of large-$N$ gauge theories with massless adjoint matter on $S^3 \times S^1$ are (meromorphic) modular forms. Our results generalize our earlier findings from  Ref.~\cite{Basar:2015xda} for pure Yang-Mills theory to theories with matter, and hold for both thermal and $(-1)^F$-twisted partition functions. Consequently, we were able to show that the confined-phase spectra of adjoint-matter gauge theories coincide with the spectra of chiral sectors of certain 2D CFTs. This means that the spectra of large-$N$ confining theories are organized by the symmetries of 2D CFTs, at least in the limit we considered.  

It is important to emphasize that our results use the large-$N$ limit in an essential way. Perhaps the simplest way to appreciate this is to recall that from start to finish, we work in finite spatial volume. (To avoid possible confusion, we note that the $S^3$ volume is always strictly finite in units of the $S^1$ size $\beta$. For most of the theories we consider, there is also a strong scale $\Lambda$, and we work in a zero-volume limit with respect to $\Lambda$, so that $R \Lambda \to 0$.)  For finite $N$ and finite volume, there is no sharp distinction between confined and deconfined phases, nor can there be any non-analyticities in the partition function.  As discussed in Ref.~\cite{Aharony:2003sx}, non-analyticities such as Hagedorn poles can appear only in systems with an infinite number of degrees of freedom. Thus, while non-analyticities can appear in infinite-volume theories at any $N$, at \emph{finite} volume non-analyticities can only arise in the infinite-$N$ limit. Consequently, a finite-$N$ thermal partition function $Z$ necessarily contains contributions from both the confined and deconfined ``phases", and is smooth as a function of $\beta$. But at small $\beta$, it is then unavoidable that the behavior of $Z$ will be that of the deconfined ``phase", and $\log Z$ will diverge as $\beta^{-3}$. In view of the general arguments we have advanced here, this implies that finite-$N$ thermal partition functions $Z$ cannot be written in terms of modular forms. Thus, within the setting we consider, modularity can only emerge at large $N$. The fact that modularity only appears at large $N$ is actually encouraging in view of our original motivation of understanding the large-$N$ spectrum  --- it means that the symmetries implied by the modularity are a consequence of the large-$N$ limit, and not purely due to the $\lambda \to 0$ limit we employed in order to perform our calculations.

%%%%%%%%%%%%
\subsection{Relation to prior work}
\label{sec:literatureComments}
%%%%%%%%%%%%

Our results are not the first concerning relations between 4D and 2D theories. It is therefore important to understand the relevance of our work within the context of previous results.

In several ways, our results resemble those of Ref.~\cite{Beem:2013sza}, where it was shown that certain special partition functions (`Schur indices') of $\mathcal{N}=2$ supersymmetric gauge theories are controlled by 2D chiral algebras and  thus have modular properties~\cite{Razamat:2012uv,Cordova:2015nma,Bourdier:2015wda,Bourdier:2015sga}, even at finite $N$. The common elements between our results and those of Ref.~\cite{Beem:2013sza} are that the 2D CFTs relevant for Ref.~\cite{Beem:2013sza} are generally non-unitary and logarithmic, as has also been the case for us. Furthermore, the 4D partition functions considered in Ref.~\cite{Beem:2013sza} coincide with chiral characters of these 2D CFTs, which also matches what we find. These points of agreement lead us to suspect that there may be important relations between our results and those of Ref.~\cite{Beem:2013sza} and other works on the modular structure of Schur indices. 

However, there are also some major differences between e.g. Ref.~\cite{Beem:2013sza} and our results. The construction employed within Ref.~\cite{Beem:2013sza} leverages supersymmetry in an essential way by noting that the only states that make non-cancelling contributions to Schur indices live on a two-dimensional plane. Once this feature of Schur indices is recognized, the appearance of a 2D chiral algebra organizing the spectrum of states contributing to these indices becomes natural.  In contrast, supersymmetry is irrelevant to our construction:  indeed our analysis applies to not only twisted partition functions but also {\it thermal}\/ partition functions, where all states contribute with the same sign and thus cannot cancel against each other. As a result,  our 4D-2D relations apply to \emph{all}\/ finite-energy states of the 4D large-$N$ theory, and not just a subset which propagates in a two-plane. Viewed from this perspective, the conceptual origin of the 2D description of our partition functions is much more mysterious than that in Ref.~\cite{Beem:2013sza}. Finally, our results apply only for large $N$, while the results of Ref.~\cite{Beem:2013sza} apply for any finite~$N$. 

%%%%%%%%%%%%%%
\subsection{Open questions}
%%%%%%%%%%%%%%
Our results suggest a large number of interesting questions:

\begin{itemize}
\item It is important to explore the connection between our results and string-theoretic expectations. 
From string theory, one might have expected that the single-particle spectrum would have a description in terms of vibrations of a string. The physics of a single string has a worldsheet CFT description. Consequently, one might have expected that the single-particle spectrum (which is just the single-trace spectrum) of a large-$N$ gauge theory would have the simplest 2D CFT description, if one were possible.  However, in contrast to this naive expectation, we have found that it is the \emph{grand-canonical}\/ partition function --- the partition function which contains contributions from all multi-trace states ---  that has a simple 2D CFT description.  Another potential issue is that a modular structure is required for the \emph{worldsheet} partition function of a string theory, and the worldsheet and spacetime partition functions do not normally coincide.  Yet one might expect that the field-theory partition function would be related to the spacetime partition function of the string theory (in a holographic way).  These issues make a stringy interpretation of our results an interesting challenge.

\item As remarked above, it is important to try to understand the meaning of $\re \tau$ on the 4D sides of our 4D-2D equivalences. Within 2D CFT chiral partition functions, turning on $\re \tau \neq 0$ corresponds to turning on a chemical potential for angular momentum on the spatial cycle of the torus. Equivalently, turning on $\re \tau$ amounts to counting states in the partition function with a twist related to their angular momentum. 

In our 4D theories, in the limit $\lambda \to 0$, the energy $E$ of a generic multiparticle states happens to coincide with their total angular momentum $J$. Both $E$ and $J$ are conserved quantities which are bounded from below. Turning on $\re \tau \neq 0$ can thus be interpreted as twisting the 4D partition function by either of these conserved quantities. Sometimes such twists coincide with standard notions.  For instance, in theories with fermions, $\tau \to \tau+1$ changes the fermion boundary conditions from periodic to anti-periodic.  While twists by $E$ or $J$ seem well defined from a statistical-mechanics perspective, it is not clear to us how to interpret such operations within a Euclidean path integral formulation of a quantum field theory.  Thus, for now, it is probably safest to view turning on $\re \tau \neq 0$ as an analytic continuation of the 4D partition function. Analytic continuation of path integrals (and hence partition functions) has recently been the focus of many works;  see, e.g., Refs.~\cite{Witten:2010cx,Basar:2013eka,Cherman:2014ofa,Behtash:2015zha,Behtash:2015loa}. Nevertheless, it would be satisfying to find a direct physical interpretation of moving along the $\re \tau$ axis in the 4D theory. 

\item A possibly related issue is to find a 4D gauge-theory interpretation of the modular images of the 4D partition functions.   It seems conceivable that more generally, the modular images of the confined-phase partition functions could be obtained by computing partition functions with background fields turned on, perhaps fields coupling to some extended operators.\footnote{We thank Chris Beem for comments on this issue.}  It is also possible that understanding the modular images of the 4D partition functions might help in understanding the meaning of $\re \tau$, because even if $\tau$ starts on the imaginary axis, modular transformations can map it to many locations within the complex plane.

\item Two-dimensional CFTs have symmetry algebras that include the Virasoro algebra.  Our 4D-2D correspondence then suggests that the symmetries of 4D confining theories should include a Virasoro symmetry acting on the spectrum.  It would be very interesting to show this explicitly within the 4D theory, and to explicitly exhibit the symmetry generators in terms of the fields of the 4D theory.

\item As suggested by the analysis of Refs.~\cite{Witten,Sundborg:2000wp}, we expect that our 4D gauge theories have an infinite tower of higher-spin conserved currents in the $\lambda \to 0$ limit. This makes it important to understand whether the 2D CFTs appearing in our 4D-2D relation also have a tower of conserved higher-spin currents, which would mean that their symmetries involve $\mathcal{W}$-algebras. 

\item It would be very interesting to extend our spectral 4D-2D equivalence to include correlation functions as well. If this turns out to be possible, a dictionary relating correlation functions in 4D and 2D would presumably shed light on the otherwise mysterious fact that the 2D CFTs we wrote down are non-unitary. 

\item It may also be important to determine whether there is a connection between the modular properties of the $\mathcal{N}=4$ SYM thermal partition function and the Yangian spectrum-generating algebra of $\mathcal{N}=4$ theory~\cite{Beisert:2010jq}. If there is such a connection, it could have important implications for understanding whether integrability of the planar spectral problem might extend to some non-supersymmetric large-$N$ theories.
 
\item Finally, perhaps the most important issue is to understand what happens to our 4D-2D equivalence away from $\lambda = 0$. If the modular structure of the partition functions generalizes in some fashion to finite $\lambda$, this would have potentially important implications for the symmetries of confining gauge theories at generic values of $R\Lambda$. To  understand whether this is possible, it may be helpful to first understand how the 4D-2D relation generalizes to correlation functions. This might then enable the development of a mapping between the finite-$\lambda$ deformation of the 4D theories and some equivalent deformation of the  2D theories. 

\end{itemize}

This list of open questions just scratches the surface of the topic of 4D-2D relations for non-supersymmetric large $N$ theories exposed by our results.  We hope that explorations of some of these issues will lead to a better understanding of confining gauge theories.

\acknowledgments
We are grateful to Chris Beem, Simon Caron-Huot, Heng-Yu Chen, David Gross, Sergei Gukov, David B. Kaplan,  Andreas Karch, Zohar Komargodski, Leonardo Rastelli, Edgar Shaghoulian, Bo Sundborg, Mithat \"Unsal, and Larry Yaffe for helpful discussions.  GB is supported by the U.S Department of Energy under Grant DE-FG02-93ER-40762, while AC is supported by U.S. Department of Energy under Grant DE-FG02-00ER-41132, and KRD is supported in part under Grant DE-FG02-13ER-41976. The research activities of KRD are also supported in part by the National Science Foundation through its employee IR/D program. The opinions and conclusions expressed herein are those of the authors, and do not represent any funding agencies.

\appendix
\section{Conventions for modular and elliptic forms}
\label{sec:AppendixConventions}
%%%%%%%%%%%%%

Our conventions\footnote{These conventions follow those of Chap.~7, Sect.~2, of Ref.~\cite{Polchinski:1998rq}.} for the Jacobi theta-functions are given by
\bea
\theta_1(z,\tau)&\equiv &-i\sum_{n\in\mathbb Z} (-1)^n     \zeta^{n+1/2} q^{(n+1/2)^2\over 2} \nonumber\\
\theta_2(z,\tau)&\equiv &\sum_{n\in\mathbb Z} \zeta^{n+1/2}  q^{(n+1/2)^2\over 2}\nonumber\\ 
\theta_3(z,\tau)&\equiv &\sum_{n\in\mathbb Z} \zeta^n q^{n^2\over 2}\nonumber\\ 
\theta_4(z,\tau)&\equiv &\sum_{n\in\mathbb Z} (-1)^n \zeta^n q^{n^2\over 2} 
\ea
where 
\begin{align}
\zeta \equiv  e^{2 \pi i z} \quad,  \quad q \equiv e^{2\pi i \tau}~. 
\end{align}
These functions transform under modular transformations $T:\tau\rightarrow\tau+1$ and $S:\tau\rightarrow -1/\tau$ 
according to
\bea
\theta_1(z,\tau+1)&=& e^{i \pi / 4 } \theta_1(z,\tau),\qquad \theta_1(z,-1/\tau)= i \sqrt{-i\tau}e^{ i \pi \tau z^2} \theta_1(-\tau z,\tau)\nonumber\\
\theta_2(z,\tau+1)&=& e^{i \pi /4 } \theta_2(z,\tau),\qquad \theta_2(z,-1/\tau)=\sqrt{-i\tau}e^{i \pi \tau z^2} \theta_4(-\tau z,\tau)\\
\theta_3(z,\tau+1)&=& \theta_4(z,\tau),\qquad\qquad \theta_3(z,-1/\tau)=\sqrt{-i\tau}e^{i \pi \tau z^2} \theta_3(-\tau z,\tau)\nonumber\\
\theta_4(z,\tau+1)&=& \theta_3(z,\tau),\qquad\qquad\theta_4(z,-1/\tau)=\sqrt{-i\tau}e^{i \pi \tau z^2} \theta_2(-\tau z,\tau)\,.\nonumber
\ea
A shorthand notation for the $\zeta=0$ special case is $\theta_i(\tau)\equiv \theta_i(0,\tau)$. The Jacobi functions have infinite-product representations given by
\begin{eqnarray}
\frac{\theta_1(z,\tau)}{2 q^{\frac{1}{8}} \sin \pi z} &=& \prod_{n=1}^\infty (1 - q^n) (1 - 2 q^n\cos \pi 2z + q^{2n}) = \prod_{n=1}^\infty (1 - q^n) (1 - q^n \zeta)(1 - q^n/\zeta) \nonumber\\
\frac{\theta_2(z,\tau)}{2 q^{\frac{1}{8}} \cos \pi z} &=& \prod_{n=1}^\infty (1 - q^n) (1 + 2 q^n\cos \pi 2z + q^{2n}) =\prod_{n=1}^\infty (1 - q^n) (1 + q^n \zeta) (1 + q^n/\zeta) \\
\theta_3(z,\tau) &=& \prod_{n=1}^\infty (1-q^n) (1+2 q^{n-\half}\cos \pi 2z+q^{2n-1}) = \prod_{n=1}^\infty (1 - q^n) (1 + q^{n-\half} \zeta) (1 + q^{n-\half}/\zeta) \nonumber\\
\theta_4(z,\tau) &=& \prod_{n=1}^\infty (1-q^n) (1-2 q^{n-\half}\cos \pi 2z+q^{2n-1}) = \prod_{n=1}^\infty (1 - q^n) (1 - q^{n-\half} \zeta) (1 - q^{n-\half}/\zeta)\,.\nonumber
\end{eqnarray}
We next define the generalized theta-function $\Th{\alpha}{\beta}(\tau)$: 
\bea
\Th{\alpha}{\beta}(\tau) ~\equiv~ \sum_{n\in\mathbb Z} e^{2\pi i n \beta} q^{(n+\alpha)^2\over 2} ~.  
\label{Th}
\ea
These functions  also have a triple-product form:
\begin{align}
\Th{\alpha}{\beta}(\tau) &=e^{i \pi \tau \alpha^2 } \prod_{n=1}^{\infty} (1-e^{2 i \pi \tau n})(1+e^{2 i\pi\tau(n-\half+\alpha)+2i\pi \beta })(1+e^{2 i\pi\tau(n-\half-\alpha)-2i\pi \beta })\nonumber\\
&=q^{\alpha^2/2}\prod_{n=1}^{\infty} (1-q^n)(1+q^{n-\half+\alpha}e^{2i\pi \beta })(1+q^{n-\half-\alpha} e^{-2i\pi \beta })~.
\label{eq:triple_product}
\end{align}
The standard Jacobi theta-functions $\theta_i(z,\tau)$ can be written in terms of $\Th{\alpha}{\beta}(\tau)$ as
\begin{align}
\theta_1(z,\tau)&=-ie^{i \pi z}\Th{\half}{z+\half}(\tau)\nonumber\\ 
\theta_2(z,\tau)&=e^{i \pi z}\Th{\half}{z}(\tau) \nonumber\\
\theta_3(z,\tau)&=\Th{0}{z}(\tau) \nonumber\\
\theta_4(z,\tau)&=\Th{0}{z+\half}(\tau)\,.
\end{align}
The generalized theta-function satisfies the identities
\begin{align}
\Th{\alpha+1}{ \beta}(\tau) &= e^{-2i\pi \beta}\Th{\alpha}{\beta}(\tau)\nonumber\\
\Th{\alpha}{ \beta+1}(\tau) &= \Th{\alpha}{\beta} (\tau) \nonumber\\
\Th{-\alpha}{- \beta}(\tau) &= \Th{\alpha}{\beta}(\tau) \label{thInv2} \, ,
\end{align}
and transforms under $T$ and $S$ as
\begin{align}
\Th{\alpha}{\beta}(\tau+1)&=e^{i\pi \alpha^2}\Th\alpha{ \beta+\alpha+\half}(\tau) \nonumber\\
\Th{\alpha}{\beta}(-1/\tau) &=\sqrt{-i\tau}\,e^{-2\pi i \alpha \beta }\Th{- \beta}\alpha(\tau) \label{eq:Strans2} \, .
\end{align}
The $T$-transformation follows straightforwardly from Eq.~\eqref{Th}.  We emphasize that these expressions are valid for arbitrary complex $\alpha$ and $\beta$, as can be verified by, e.g., deriving the $S$-transformations using the Poisson summation formula. 

Finally, the Dedekind eta-function is defined as
\bea
\eta(\tau)~\equiv~ q^{\frac{1}{24}}\prod_{n=1}^\infty(1-q^n)~.
\ea
This transforms as
\begin{align}
\eta(\tau+1) &= e^{i \pi/ 12} \eta(\tau) \nonumber\\
 \eta(-1/\tau) &=\sqrt{-i\tau}\,\eta(\tau)\,
\end{align}
and exhibits the double-argument relations
\begin{align}
\eta(2\tau) &={\eta^2(\tau)\over \left(\theta_4(\tau) \theta_3(\tau)\right)^{\half}}={\eta^2(\tau)\over\left(\Th{0}{\half}(\tau)\Th{0}{0}(\tau)\right)^{\half}} = {1\over \sqrt{2}} \left[\Th{\half}{0}(\tau) \eta(\tau)\right]^{\half}~.
\end{align}

%%%%%%%%%%%%%%%%%
\section{Roots for generic $n_s, n_f$}
\label{sec:RootAppendix}
%%%%%%%%%%%%%%%%%
To find expressions for the roots of $P(q)$ for generic $n_s, n_f$, we write
\begin{align}
P(q) = \prod_{i=1}^{3} (1 + R_i q +q^2)
\end{align}
and look for $R_i = r_i + r_i^{-1}$ such that 
\begin{align}
P(q) = 1 - 3 q^2 - n_s q^2 + 4 n_f q^3 - 3 q^4 - n_s q^4 + q^6 .
\end{align} 
Matching powers of $q$ and solving the resulting set of three equations yields
\begin{align}
R_1 &= -\frac{B+X^{2/3}}{\sqrt{3} \sqrt[3]{X}} \\
R_2 &=\frac{\sqrt{3} \sqrt[3]{X} \left(B+X^{2/3}\right) +3 \sqrt{2 A X+B^3-B^2 X^{2/3}+2 B X^{4/3}}}{6 X^{2/3}}\\
R_3 &=\frac{\sqrt{3} \sqrt[3]{X} \left(B+X^{2/3}\right) - 3 \sqrt{2 A X+B^3-B^2 X^{2/3}+2 B X^{4/3}}}{6 X^{2/3}}
\end{align}
where
\begin{align}
A &= 6 \sqrt{3} n_f\,,\qquad
B =6+n_s \,, \qquad
X = \sqrt{A^2-B^3}-A~.
\end{align}

These expressions are valid for any $n_s, n_f$. However, as mentioned in the main text, for certain select values of $n_s, n_f$, there are dramatic simplifications, with some roots becoming~$1$. This is the fundamental reason why SUSY theories have different, and slightly simpler, modular structures than their non-SUSY cousins. For instance, for the theory with $\mathcal{N}=4$ SYM matter content, we find
\begin{align}
n_s=6, n_f=4 \; : \qquad (R_1,R_2, R_3) = (-2,-2,4),
\end{align}
so that
\begin{align}
P(q)\big|_{\mathcal{N}=4} = (1-2q+q^2)^2(1+4q +q^2) = (1-q)^4(1+4q +q^2).
\end{align}

%%%%%%%%%%%%%%%%
\section{Simplifications at $n_f = 0$}
\label{sec:BosonicSimplifcations}
%%%%%%%%%%%
In this appendix, we show how the formulas derived in Sects.~\ref{sec:twistedPFs} and \ref{sec:thermalPFs}---formulas which hold for any non-supersymmetric choice of $n_s$ and $n_f$, including $n_f=0$---match the seemingly different expressions we obtained in Sect.~\ref{sec:Bosonic} for theories with purely bosonic matter content. 

We begin by noting how the modular-form expressions, derived for generic $n_f$ and $n_s$, simplify when $n_f = 0$. First, we rewrite the defining polynomial for the purely bosonic theories in terms of the variable $Q^2 \equiv q$:
\begin{align}
P(q^n) &= (1+q^n)(1-(4+n_s)q^n + q^{2n}) \nonumber \\ 
&= (1 + q^n) (q^n - z(A))\big|_{A=+1} (q^n - z(A) )\big|_{A=-1} 
\label{bosonPoly} 
\end{align}
where
\begin{align}
\quad z(A) = \left(2 + \frac{n_s}{2}\right) +A \sqrt{\left(2+\frac{n_s}{2} \right)^2-1}~. 
\label{bosonQz}
\end{align} 
In Eq.~(\ref{bosonQz}) we see how the $Q$-variable polynomial factorizes;  indeed we have $z(+1) z(-1) = 1$. In terms of $q$, we then obtain
\begin{align} 
P_{\rm boson}(Q) 
&= (Q+i)(Q-i) \prod_{A = \pm1} \left(Q + i \sqrt{z(A)} \right) \left(Q+\frac{1}{i \sqrt{z(A)}}\right) \nonumber\\
&= (Q-i)(Q+i) \prod_{A = \pm1} \left(Q - i \sqrt{z(A)} \right)  \left(Q -\frac{1}{i \sqrt{z(A)}}\right) ~. 
\label{eq:zsYMth}
\end{align}
Note that the two lines in Eq.~(\ref{eq:zsYMth}) differ by sign choices but nevertheless multiply out to the same expression. This sign ambiguity is related to the ambiguity in extracting a sign for $Q$ from $q$, given that $q = (-Q)^2 = (+Q)^2$. 

Given these observations, we can rewrite the large-$N$ partition function for purely-bosonic gauge theories in 
a form which resembles the partition functions of gauge theories with fermionic matter:
\begin{align}
Z_{\rm YM}(\tau) &
= \prod_{n=1}^{\infty} \frac{(1-Q^{2n})^3}{(1+Q^{2n})(1-(4+n_s)Q^{2n} + Q^{4n})} 
= \prod_{\alpha = 1}^3 \prod_{n=1}^{\infty} \frac{(1-Q^{2n})}{(1+Q^{n} z_{\alpha})(1+Q^{n}/z_{\alpha})} \nonumber \\& 
= \prod_{\alpha = 1}^3 \prod_{n=1}^{\infty} 
\frac{1}{(1+Q^{2n} z_{\alpha})(1+Q^{2n}/z_{\alpha})}
\frac{(1-Q^{n})^3}{(1-Q^{n})^2}
\frac{1}{(1+Q^{2n-1} z_{\alpha})(1+Q^{2n-1}/z_{\alpha})} \nonumber\\
& \propto \prod_{\alpha = 1}^3 \frac{1}{\theta_2(b_{\alpha},\tau)}~ \eta(\tau)^3 ~ \frac{1}{\theta_3(b_{\alpha},\tau)}
\propto \prod_{\alpha = 1}^3 \frac{\eta(\tau)^3}{\,\, \Th{1/2}{b_{\alpha}}(\tau) \,\,\,\, \Th{0}{b_{\alpha}}(\tau) \,\, } \label{eq:thYM} \, .
\end{align}
It is important to note that the sign ambiguity for the $z_{\alpha}$ in the above expressions leads to an ambiguity in the real part of the lower characteristic of $\Th{0}{b_{\alpha}}(\tau)$, since $\Th{0}{b_{\alpha}}(\tau) \sim \Th{0}{b_{\alpha}+\frac{1}{2}}(\tau)$. This only occurs for purely imaginary roots of the defining polynomial, and pure-imaginary roots are unique to $n_f = 0$. It is precisely this feature which allows the apparently dissimilar expressions for $\tilde{Z}(n_f,n_s)$ and $Z(n_f, n_s)$ to match when $n_f = 0$.

Equipped with the result in Eq.~\eqref{bosonQz}, we can now find the specific $\{ b_{\alpha} \}$ which enter into Eq.~\eqref{eq:thYM} and moreover verify analytically that the sum $\sum_{\alpha} (b_{\alpha})^2$ is real. Had this not been real, the ``phase-factors'' in Eq.~\eqref{eq:TmnTActionGeneral} would have had non-unit modulus. Unit modulus phase-factors are crucially tied to the convergence of the modular orbits. As we shall see, it is simplest to show that these phase factors are indeed pure phases for the special case of $n_s = 0$. Proving these reality conditions for general $n_s \neq 0$ will then be relatively straightforward. 

For $n_s = 0$, the $\lbrace z_{\alpha}\rbrace $ which enter into Eq.~\eqref{eq:thYM} are simply given by
\begin{eqnarray}
z_{1} &=& i~ \nonumber\\  
z_{2} &=& i \left( 2 + \sqrt{3} \right)^{\frac{1}{2}}~ \nonumber\\
z_{3} &=& i \left( \frac{1}{2 + \sqrt{3}} \right)^{\frac{1}{2}} = i \left( 2 - \sqrt{3} \right)^{\frac{1}{2}} ~.
\label{YMroot} 
\end{eqnarray}
The relation $z_{\alpha} = e^{2\pi i b_{\alpha}}$ then allows us to  solve directly for the $\{ b_{\alpha} \}$:
\begin{eqnarray}
b_{1} &=& \frac{1}{2\pi i} \log(i) = \frac{1}{4} \nonumber\\
b_{2} &=& \frac{1}{2\pi i} \left( \log(i) + \log\left( 2 + \sqrt{3} \right)^{\frac{1}{2}} \right) = \frac{1}{4} + i B 
\nonumber \\
b_{3} &=& \frac{1}{2\pi i} \left( \log(i) - \log\left( 2 + \sqrt{3} \right)^{\frac{1}{2}} \right) = \frac{1}{4} - i B ~ . 
\label{b3}
\end{eqnarray}
From this it follows that
\begin{align}
\sum_{\alpha} (b_{\alpha})^2 \bigg|_{\rm YM} =
\left(\frac{1}{4} \right)^2 +
\left(\frac{1}{4} + i B \right)^2 +
\left(\frac{1}{4} - i B \right)^2
 = \frac{3}{16} - 2 B^2 \, . \label{eq:bReal}
\end{align}
We observe that the reality of this sum is guaranteed simply because the two non-trivial complex characteristics are conjugate to each other. This conjugate nature, ensuring the reality of the expression in Eq.~\eqref{eq:bReal}, is fundamentally due to the alternating signs on the square roots present in the initial defining polynomials in Eqs.~\eqref{bosonPoly} and~\eqref{bosonQz}.

Generalizing the reality condition in Eq.~\eqref{eq:bReal} for $n_s \neq 0$ is straightforward. Substituting
\begin{align}
2 + \sqrt{3} ~\longrightarrow~\left( 2 + \frac{n_s}{2} \right) + \sqrt{\left( 2 + \frac{n_s}{2} \right)^2 -1} 
\end{align}
again yields $b_{1}(n_s) = i$ and $b_{2}(n_s) = \overline{b_{3}(n_s)}$. Hence $\sum_{\alpha =1}^3 (b_\alpha)^2 \in \mathbb{R}$ for all $n_s$.

%%%%%%%%%%%%%%%%%%%%%%%%%%%%%%%%%%%%%%%%%%%%%%%%%%%
\section{Alternate definitions of $\tau$ and extra simplifications for ${\cal N} = 4$ SYM}
\label{sec:OtherModular}
%%%%%%%%%%%%%%%%%%%%%%%%%%%%%%%%%%%%%%%%%%%%%%%%%%%

In the main body of the paper we defined the parameter $\tau$ by analytic continuation from $\beta/R$, where $R$ is the radius of the three-sphere on which we are compactifying our 4D gauge theories and $\beta$ is the circumference of the thermal circle. Specifically, we analytically continued $\beta/R \to  -2\pi i \tau$  with $\tau\in\mathbb{C}$, whereupon we see that
\begin{align}
\im \tau = \frac{1}{2 \pi}\frac{\beta}{R}~ 
\label{eq:STDchoice}
\end{align}
and $q\equiv e^{-\beta/R} \to  e^{2\pi i \tau}$. However, we have not found a satisfying physical interpretation of $\re \tau$ within the 4D gauge theory. In this appendix, we explore the consequences of the fact that other definitions of $\tau$ are also possible. Our hope is that these remarks might be helpful for future studies which might seek to explore the meaning of $\re \tau$ for 4D gauge theories.

Let us first recall the consequences of this definition of $\tau$. With this definition of $\tau$, the modular $T$-transformation $\tau \to \tau+1$ has the effect of changing the fermion boundary conditions in the Euclidean path-integral language, or equivalently has the effect of inserting $(-1)^F$ into the partition function in Hamiltonian language. To see this, recall that in free theories on $S^3_R \times \mathbb{R}$ bosonic states have energies $\omega_{n,B} = n/R$ while fermions have energies $\omega_{n,F} = (n+\half)/R$. Consequently, when bosonic and fermionic states appear in partition functions, they are associated with factors of $q^n$ and $q^{n+\half}$ respectively. Thus, under $T$, bosonic energy contributions to partition functions are unaffected, while fermionic contributions are multiplied by a factor of $(-1)$. This is precisely the effect of inserting a $(-1)^F$ operator into the trace over Hilbert space defining a partition function.

We could instead define a modular parameter $\tau^{[x]}$ by analytically continuing  $\beta/R \to - 2\pi i \, x\, \tau^{[x]}$ for any $x\in \mathbb{C}$ with $\re[x] >0$, so that
\begin{equation}
q \equiv  e^{-\frac{\beta}{R}} \longrightarrow e^{2\pi i  x \tau^{[x]}} ~. 
\label{eq:tauX}
\end{equation}
To see the effect of this, let us first consider the action of the modular $T$-transformation $T: \tau^{[x]} \to \tau^{[x]} +1$ on the partition function for the bosonic and fermionic states, which is determined by the action of $T$ on $q^n$ and $q^{n+1/2}$ respectively:
\begin{align}
T : \qquad   &  q^n  \longrightarrow e^{2 \pi i n x } q^{n} ~~ ,  \quad 
q^{n+1/2} \longrightarrow e^{2\pi i (n+1/2) x} q^{n+1/2}. 
\end{align}
For integer $x$, the bosonic and fermionic Boltzmann factors are mapped into themselves up to an overall sign of $\pm 1$, while for non-integer values of $x$ they accrue non-trivial phases.  Integer values of $x$ are clearly rather special, in that when $x \in \mathbb{Z}$ the modular $T$-transformation has a simple action. In the body of the paper we took $x = 1$, and in this case the effect of the $T$-transformation is to flip the sign of the fermionic Boltzmann factors.  So acting with $T$ amounts to a change in the fermion boundary conditions in the Euclidean path-integral formulation of the theory when $x=1$. 

In this appendix, by contrast, we explore the consequences of choosing the $x = 2$ proportionality factor, so that $\im[\tau^{[2]}] = \frac{1}{2\pi} \frac{\beta}{2R}$.  With this definition of the modular parameter, modular transformations do \emph{not} change boundary conditions of either the fermions or the bosons on $S^1$. As such, the modular orbits of both fermionic and bosonic large-$N$ gauge theories are significantly simpler. Indeed, the reason why the modular orbits with $\tau^{[2]} = \frac{1}{2\pi i} \frac{\beta}{2R}$ are ultimately simpler then those with $\tau^{[1]} = \frac{1}{2\pi i} \frac{\beta}{R}$ is because the modular group associated with the former variable is a subgroup of that associated with the latter.  

For the rest of this section, we use $\tau^{[2]}$ as the modular parameter, so that
\begin{align}
Q &= e^{-\beta/2R} = e^{2 \pi i \tau^{[2]}} ~ , \quad q=Q^2  =e^{-\beta/R} = e^{4 \pi i \tau^{[2]}} ~.
\label{eq:Qd2}
\end{align}
In this notation, the twisted partition function for YM coupled to a generic number of adjoint scalars and adjoint fermions in Eq.~\eqref{eq:twistedPF} evaluates to
\begin{align}
\tilde{Z}\big(\tau^{[2]}\big) = \prod_{\alpha = 1}^{3} \prod_{n=1}^{\infty} \frac{(1-Q^{2n})}{(1+Q^n z_{\alpha})(1+Q^n/z_{\alpha})} = \prod_{\alpha = 1}^{3} \left( \frac{\eta\big(\tau^{[2]}\big)}{\theta_2\big(b_{\alpha},\tau^{[2]}\big)} \big\{\eta\big(\tau^{[2]}\big) \theta_2\big(0,\tau^{[2]}\big) \big\}^{\frac{1}{2}} \right) \label{eq:twisteD} \, ,
\end{align}
which again has a modular structure. Here $z_{\alpha}$ and $b_{\alpha}$ have the same definitions as in the body of the paper and in Appendix~\ref{sec:RootAppendix}, and implicitly we have assumed that $z_{\alpha} \neq -1$. Thermal partition functions can be obtained from the above by sending $\tau^{[2]} \to \tau^{[2]} +\frac{1}{2}$, which is consistent with the claim that the modular $T$ transformation do not change the fermion boundary conditions.  This feature turns out to make the modular orbit of $\tilde{Z}(\tau^{[2]})$ simpler than the modular orbit of $\tilde{Z}(\tau^{[1]})$ discussed in the body of this paper. 

To make the essential points in the simplest context, consider the modular orbit for the large-$N$ limit of $\mathcal{N}=4$ SYM theory. When expressed in terms of $Q = e^{-\beta/2R}$, the starting ``seed'' twisted partition function takes the form
\begin{align}
\tilde{Z}_{{\cal N} = 4}\big(\tau^{[2]}\big) = \frac{\cos(\pi b) }{\sqrt{2}}\frac{1}{\eta\big(\tau^{[2]}\big)} \left(\frac{\Th{1/2}{0}\big(\tau^{[2]}\big)}{\eta\big(\tau^{[2]}\big)}\right)^{3/2} \left( 
\frac{\eta\big(\tau^{[2]}\big)}{\Th{\frac{1}{2}}{b}\big(\tau^{[2]}\big)} \right)~. 
\label{eq:SYMd1}
\end{align}
It is easily seen that this ``seed'' term falls into the more general class of terms given by 
\begin{align}
\tilde{T}^{{\cal N} =4}_{m,n}\big(\tau^{[2]}\big) = \frac{\cos(\pi b) }{\sqrt{2}}\frac{1}{\eta\big(\tau^{[2]}\big)} \left(\frac{\Th{P(m)/2}{P(n)/2}\big(\tau^{[2]}\big)}{\eta\big(\tau^{[2]}\big)}\right)^{3/2} \left( 
\frac{\eta\big(\tau^{[2]}\big)}{e^{\pi i P(m) \cdot n \cdot b}\,\Th{m b + \frac{P(m)}{2}}{n b + \frac{P(n)}{2}}\big(\tau^{[2]}\big)} \right) 
\, \label{eq:SYMd2}
\end{align}
where $m,n$ are again coprime integers. Manipulations isomorphic to those in Sects.~\ref{sec:BosonicModularCompletion} and~\ref{sec:GenericModularCompletion} establish that the $S$- and $T$-transformations indeed act within the set of functions defined in Eq.~\eqref{eq:SYMd2} and that the modular orbits of $\tilde{Z}_{{\cal N} = 4}(\tau)$ in Eq.~\eqref{eq:SYMd2} map surjectively into the set of coprime pairs.  As a result, the twisted seed term has a modular completion given by
\begin{align}
\tilde{Z}^{\mathcal{N}=4}_{\rm modular}\big(\tau^{[2]}\big) = \big(\im \tau^{[2]}\big)^{-1/2}\, 
 \sum_{ \substack{ m, n \in \mathbb{Z} \\ m\perp n } }\, \big|T_{m,n}\big(\tau^{[2]}\big)\big|^2 \, .
\label{eq:N4TwistedOrbit}
\end{align}
Again, as discussed in Sect.~\ref{sec:BosonicModularCompletion}, we immediately infer that this sum represents a 2D CFT with a collection of primary fields that have a continuum of effective conformal dimensions $h^{\rm (eff)}_k$. 

It amusing to note that neither the twisted partition function seed term \emph{nor} its modular completion have any Hagedorn poles for ``physical'' temperatures, i.e., $1/\beta = T \in [0,\infty) \subset \mathbb{R}$. The absence of Hagedorn poles in the seed term is due to intricate cancellations between bosonic and fermionic states at \emph{different}\/ levels within the twisted partition function. This observation was originally made in Ref.~\cite{Basar:2014jua}, and appears to be the first known field-theoretic incarnation of certain string-theoretic observations pertaining to boson/fermion cancellations~\cite{Kutasov:1990sv,Dienes:1994np,Dienes:1995pm} and misaligned supersymmetry~\cite{Dienes:1994np,Dienes:1995pm,Dienes:2001se}. The fact that it extends to the modular completion --- which is modular-invariant by construction --- may be important for understanding the links between our large-$N$ gauge-theory construction and string theory.

\bibliographystyle{apsrev4-1}
\bibliography{super_susy} 

\end{document}